\documentclass[%
 aip,
 apl,%
 amsmath,amssymb,
 preprint,%
]{revtex4-1}
\usepackage{hyperref}
\usepackage{graphicx}
\usepackage{epstopdf}
\usepackage{dcolumn}
\usepackage{bm}
\usepackage[caption=false]{subfig}
\usepackage{floatrow}

\def\L{\mathrm{L}}
\def\X{\mathrm{X}}
\def\K{\mathrm{K}}

\def\v{\mathrm{v}}
\def\c{\mathrm{c}}
\def\e{\mathrm{e}}
\def\h{\mathrm{h}}
\def\B{\mathrm{B}}

\begin{document}


\title[the title]{Large photoluminescence enhancement by an out-of-plane magnetic field in exfoliated WS$_2$ flakes}

\author{Sibai Sun}
\author{Jianchen Dang}
\author{Xin Xie}
\author{Yang Yu}
\author{Longlong Yang}
\author{Shan Xiao}
\author{Shiyao Wu}
\author{Kai Peng}
\author{Feilong Song}
\affiliation{Beijing National Laboratory for Condensed Matter Physics, Institute of Physics, Chinese Academy of Sciences, Beijing 100190, China}
\affiliation{CAS Center for Excellence in Topological Quantum Computation and School of Physical Sciences, University of Chinese Academy of Sciences, Beijing 100049, China}%
\author{Yunuan Wang}
\affiliation{Beijing National Laboratory for Condensed Matter Physics, Institute of Physics, Chinese Academy of Sciences, Beijing 100190, China}
\affiliation{Key Laboratory of Luminescence and Optical Information, Ministry of Education, Beijing Jiaotong University, Beijing 100044, China}

\author{Jingnan Yang}
\author{Chenjiang Qian}
\author{Zhanchun Zuo}
\affiliation{Beijing National Laboratory for Condensed Matter Physics, Institute of Physics, Chinese Academy of Sciences, Beijing 100190, China}
\affiliation{CAS Center for Excellence in Topological Quantum Computation and School of Physical Sciences, University of Chinese Academy of Sciences, Beijing 100049, China}%
\author{Xiulai Xu}%
\email{xlxu@iphy.ac.cn}
\affiliation{Beijing National Laboratory for Condensed Matter Physics, Institute of Physics, Chinese Academy of Sciences, Beijing 100190, China}
\affiliation{CAS Center for Excellence in Topological Quantum Computation and School of Physical Sciences, University of Chinese Academy of Sciences, Beijing 100049, China}
\affiliation{Songshan Lake Materials Laboratory, Dongguan, Guangdong 523808, China}

\date{\today}

\begin{abstract}
We report an out-of-plane magnetic field induced large photoluminescence enhancement in WS${}_2$ flakes at $4$ K, in contrast to the photoluminescence enhancement provided by in-plane field in general. Two mechanisms for the enhancement are proposed. One is a larger overlap of electron and hole caused by the magnetic field induced confinement. The other is that the energy difference between $\Lambda$ and K valleys is reduced by magnetic field, and thus enhancing the corresponding indirect-transition trions. Meanwhile, the Land\'e g factor of the trion is measured as $-0.8$, whose absolute value is much smaller than normal exciton, which is around $|-4|$. A model for the trion g factor is presented, confirming that the smaller absolute value of Land\'e g factor is a behavior of this $\Lambda$-K trion. By extending the valley space, we believe this work provides a further understanding of the valleytronics in monolayer transition metal dichalcogenides.
\end{abstract}

\maketitle

\floatsetup[figure]{style=plain}

\section{\label{sec:intro}Introduction}
Recently the optical properties with valley feature of transition metal dichalcogenides (TMDs) have been investigated intensively.\cite{https://www.doi.org/10.1103/PhysRevLett.108.196802,PhysRevLett.105.136805, doi:10.1021/nl903868w, doi:10.1021/nn305275h, doi:10.1021/nn5021538} Especially, the magneto-optical properties have raised great attentions since magneto-photoluminescence spectroscopy is a promising tool to investigate spin and valley properties of excitons.\cite{ISI:000349934700017,ISI:000367748600015,https://arxiv.xilesou.top/pdf/2002.11646.pdf,ISI:000412635500004,10.7498/aps.67.20180615} Normally, the valley and spin information is locked by the selection rule. By applying a magnetic field, the degeneracy of the spin is lifted, and the valley information will change accordingly. Due to the spin-orbit couplings of conduction band and valence band have opposite sign \cite{https://doi.org/10.1515/nanoph-2016-0165,https://www.doi.org/10.1103/PhysRevB.88.085433}, the lower energy state does not emit photons because of the selection rule for darkish materials. \cite{https://www.doi.org/10.1088/2053-1583/aa5521} In-plane magnetic field induces tunneling between two spin status, and thus unlocking the restriction between spin and valley information, making the dark excitons bright. Therefore, the photoluminescence (PL) can be enhanced by applying an in-plane magnetic field in monolayer darkish materials. \cite{https://www.doi.org/10.1088/2053-1583/aa5521,https://www.doi.org/10.1038/NNANO.2017.105} However, different from in-plane magnetic field, out-of-plane magnetic field induced PL enhancement in monolayer TMDs has yet to be explored.

In addition to the external field modulating the PL, the valley features might also affect the PL intensity. Recently, the intervalley excitons including K, $\Lambda$ valleys (middle point between $\Gamma$ and K, sometimes also called Q valley)\footnote{Q and $\Lambda$ points in the Brillouin zone have common $x,y$ but different $z$. For monolayer 2D materials, Q and $\Lambda$ can be treated as the same.} in the conduction band and K, $\Gamma$ valleys in the valence band has been investigated.\cite{lindlau2017identifying} The properties of these valleys can be quite different from K valleys, such as orbital magnetic momenta, spin status of eigenstates, energy shifts by strain and effective masses of carriers.\cite{https://doi.org/10.1021/nl501638a,https://doi.org/10.1007/s12274-015-0762-6,http://dx.doi.org/10.1063/1.4869142} The differences provide an opportunity to open a new field of information processing with considering the valley freedom.

In this work, we report an observation of strong PL enhancement by out-of-plane magnetic field at cryogenic temperature. Two mechanisms for the enhancement are provided. One is the increase of wavefunction overlapping between electron and hole. The other is the enhanced indirect transitions between K and $\Lambda$ valleys in WS${}_2$ flakes, which results from the decrease of the energy difference of K and $\Lambda$ valleys in out-of-plane magnetic field, accompanying with a small absolute Land\'e g factor value. 



\section{\label{sec:experiment}Experimental results and Discussions}
\begin{figure}[ht]%
\centering
\includegraphics[scale=0.7]{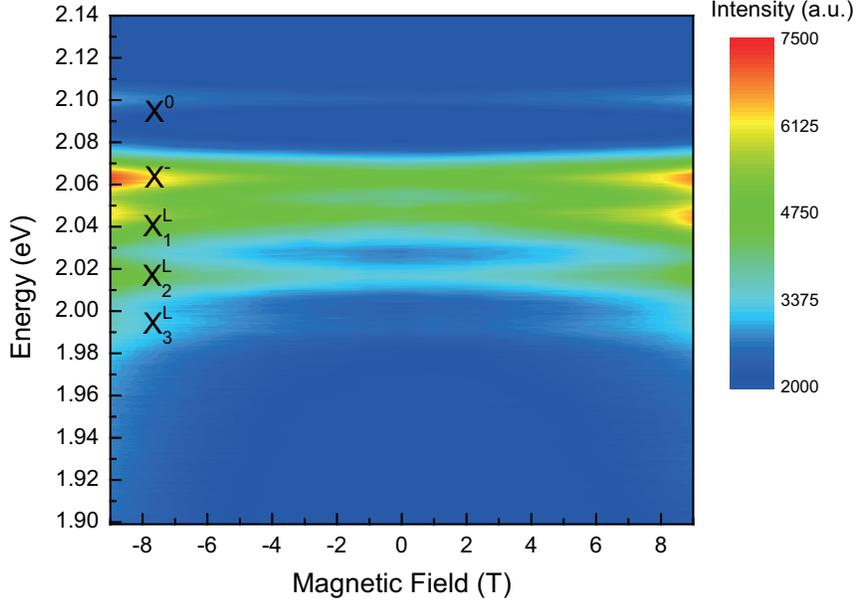}
\caption[]{\label{fig:contour}%
\raggedright
Contour plot of PL spectra from a WS$_2$ flake with a vertical magnetic field from $-9$ T to $9$ T at $4.2$ K. When the magnetic field increases, the neutral exciton peak(X${}^0$), trion peak (X${}^-$) and defect-bound excitons (X${}_{1,2,3}^{\L}$) are enhanced significantly.
}%
\end{figure}

\begin{figure}[ht]%
\centering
\includegraphics[scale=0.5]{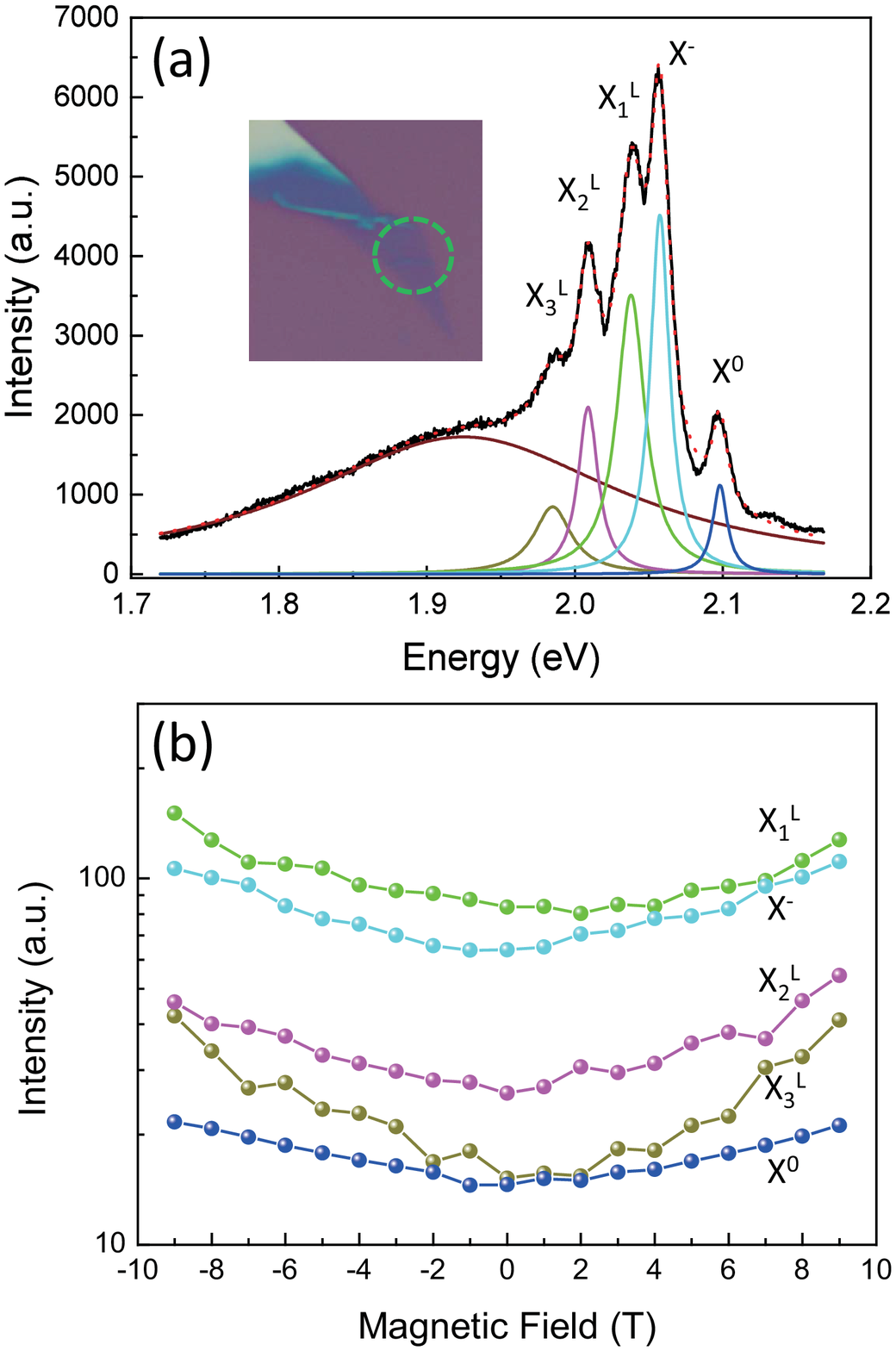}
\caption[]{\label{fig:enhancement}%
\raggedright
(a) Photoluminescence spectrum from WS$_2$ flake with magnetic field of $9$ T at $4.2$ K. The black line is the PL spectrum, and the color lines are multi-peak fitted curves. The optical microscope image is shown in the inset. (b) Peak integration intensities of X${}^0$, X${}^-$ and X${}_{1,2,3}^{\L}$ in different magnetic fields. At 9 T, the integration PL intensities of the neutral exciton (X${}^0$), trion peak (X${}^-$) and defect-bound excitons (X${}_{1,2,3}^{\L}$) are enhanced by 47\%, 70\%, 67\%, 93\% and 174\%, respectively.
}%
\end{figure}

The WS${}_2$ flakes are exfoliated from bulk materials and transferred to Si/SiO${}_2$ substrate as shown in the inset of Fig. \ref{fig:enhancement}(a). The PL spectra from WS$_2$ flake at different magnetic fields are shown in Fig. \ref{fig:contour} at 4.2 K with an excitation laser at 532 nm. The exciton peak (marked as X${}^0$) at 2.10 eV, the negative trion peaks (marked as X${}^-$) at 2.06 eV and defect-bound excitons (marked as X${}_{j}^{\L}, j=1,2,3$) at lower energy are identified.\cite{https://doi.org/10.1002/pssr.201510224} The charged trion in monolayer WS${}_2$ has been confirmed to be negative by electric tuning.\cite{http://dx.doi.org/10.1103/PhysRevLett.115.126802,https://doi.org/10.1038/srep09218} And by measuring the variation of PL intensity logarithm with different pumping power, the contribution of bi-exciton is excluded \cite{Barbone2018}. Since the direction of magnetic field is out-of-plane, the contribution of dark exciton is excluded.\cite{10.1088/2053-1583/ab2cf7} It can be seen that the intensities of PL peaks increase with increasing magnetic field. In order to obtain the intensity increase in detail, the PL spectra are fitted with multi-peak non-linear least-squares curve-fitting python package in Lorentz shape as shown in Fig. \ref{fig:enhancement} (a). When the magnetic field is increased to 9 T, the integration PL intensities of the neutral exciton (X${}^0$), trion peak (X${}^-$) and defect-bound excitons (X${}_{1,2,3}^L$) are enhanced by 47\%, 70\%, 67\%, 93\% and 174\% respectively, as shown in Fig. \ref{fig:enhancement} (b). The origin of the broad peak at 1.9 eV is still not clear, which does not show an clear enhancement with magnetic field.


The enhancement of emission with an applied magnetic field has been reported in other low dimensional semiconductor materials, such as quantum dot and quantum well systems in III-V compound semiconductors.\cite{http://link.springer.com/article/10.1007/s12274-015-0910-z,https://www.doi.org/10.1063/1.4948330,Tang27804} The magnetic field transforms the wavefunction of electron and hole and reduces wavefunction extension. And due to the large difference of effective masses of electrons and holes\cite{https://www.doi.org/10.1103/PhysRevB.43.4152}, or difference of bandstructures in heterostructure like core-shell quantum dots\cite{https://www.doi.org/10.1021/ja0361749}, the electrons and holes are spatially separated. The wavefunction extension modulation could increase the overlap of their wavefunction distribution, then influence the recombination rate, and thus resulting in a PL intensity enhancement.

Generally, in the WS${}_2$, there are neither such a large effective masses difference, nor spacial difference like heterostructure. But due to the impurities or defects, spacial difference will also be induced. For example, impurities capture carriers and form charged centers. Then excitons interact with these charged centers, leading to wavefunction radius modification by Comloub interaction. The opposite charge between electron and hole make the modification opposite, and thus induce the spacial difference. For Wannier-like excitons, the typical length scale of ground state is $a=\frac{\epsilon_r m_0}{m_{\mathrm{eff}}}a_{\B}$\cite{https://doi.org/10.1016/0038-1098(86)90573-9}, where $\epsilon_r$ is the relative dielectric constant, $m_0$ is the electron mass in vacuum, $m_{\mathrm{eff}}$ is the effective reduced mass for exciton, and $a_{\B}$ is the Bohr radius of hydrogen atom. For WS${}_2$, the $m_{\mathrm{eff}}\approx 0.5m_0$ both for electrons and holes, and relative dielectric constant $\epsilon_{\mathrm{r}}\approx 10$, \cite{http://dx.doi.org/10.1103/PhysRevB.90.205422} therefore $a\approx 1$ nm. The length scale of carrier with magnetic field can be expressed by gyroradius $r(B)=\sqrt{\frac{\hbar}{eB}}$. With magnetic field of 9 T, it is around $8$ nm. The influence of magnetic field can not be neglected for defect-bound exciton. When magnetic field is applied, the wavefunction shape of trapped carriers will be shrinking, making electrons and holes more likely to recombine.

\begin{figure}[ht]%
\centering
\includegraphics[scale=0.5]{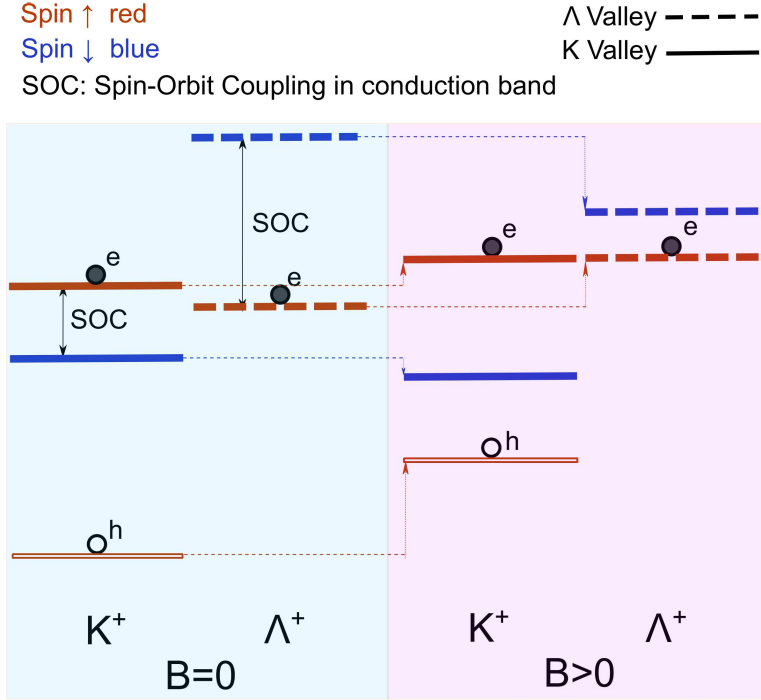}
\caption[]{\label{fig:cartoon}%
\raggedright
Band structures without (left, sky blue background) and with a magnetic field (right, violet background). Red lines represent bands whose spin eigenstate are up, and blue lines for down. The $\Lambda$-K trion contains a hole and an electron in K valley, and another electron in $\Lambda$ valley. Due to the difference of orbital terms between $\Lambda$ valley and K valley, the band shifts differently with magnetic field.
}%
\end{figure}

Meanwhile, the PL enhancement could be introduced by the reduction of the energy difference between K and $\Lambda$ valleys. As shown in Fig. \ref{fig:cartoon}, without magnetic field, the spin-up energy level of $\Lambda^{+}$ valley is lower than that of K${}^{+}$, and the spin-down energy level of $\Lambda^{-}$ valley is lower than K${}^{-}$. Here, we take spin-up situation as an example. Once magnetic field is applied on these flakes, the $\Lambda$ valley shifting is larger than the K valley shifting, which will be explained later in Eq. (\ref{eq:movement}). As a result, the energy difference between K and $\Lambda$ is reduced, as shown in Fig. \ref{fig:cartoon}. Thus the intervalley scattering between $\Lambda$ electron and K electron occurs more frequently, which can be derived from models of Raman spectrum analysis.\cite{https://www.nature.com/articles/ncomms14670} This scattering thus influences the occupation status of electrons and making the 
PL more bright. If the magnetic field is positive, the $\sigma^+$ component will be enhanced. And if it is negative, the $\sigma^-$ component will be enhanced. Therefore, related PL peaks are enhanced.

The influence of $\Lambda$ valley not only affects the PL intensity, but also the Land\'e g factors. The g factors for conduction band and valence band at each valley are expressed as
\begin{equation}
g_{b}^{V^\pm,s}=\frac{E_{b}^{V^\pm,s}(B)-E_{b}^{V^\pm,s}(0)}{\mu_{\B} B}\\
\end{equation}
where $B$ is the out-of-plane magnetic field, $b=\{\c,\v\}$ denotes conduction and valence band respectively, $V^\pm=\{\K^+,\K^-,\Lambda^+,\Lambda^-\}$ denotes the valleys, and $s=\{\uparrow,\downarrow\}$ denotes spin. For normal excitons, the Land\'e g factor then can be expressed by $g=g_{\c}^{V,\uparrow}-g_{\v}^{V,\uparrow}-g_{\c}^{V,\downarrow}+g_{\v}^{V,\downarrow}$. The linear energy level shift caused by magnetic field can be contributed by three parts: spin, orbital and valley components:\cite{https://doi.org/10.1088/2053-1583/aae14b,https://doi.org/10.1038/s41699-020-0136-0,ISI:000257289500096,ISI:000348576100021} $g_{b}^{V,s}=g_{b}^{V,s}(\mathrm{spin}) +  g_{b}^{V,s}(\mathrm{orbit}) + g_{b}^{V,s}(\mathrm{valley})$. For neutral exciton in TMDs, the spins of electron and hole are in the same direction, the spin term is zero $g_{\c}^{V,s}(\mathrm{spin})-g_{\v}^{V,s}(\mathrm{spin})=(\pm\frac{1}{2})-(\pm\frac{1}{2})=0$.\cite{https://www.doi.org/10.1038/ncomms1882} The orbital terms are mainly from the angular momentum azimuthal component of electron state around transition metal atom, $g_{b}^{V,s}(\mathrm{orbit})=\langle \hat{L}_z \rangle/\hbar$. For the conduction band at K point, the electron mainly possesses $|\K_\c^{\pm}\rangle=|d_{z^2}\rangle$ orbit and thus the azimuthal component is $\langle \K_\c^{\pm}| \hat{L}_z |\K_\c^{\pm}\rangle=0$. For the valence band at K point, the state mainly possesses $|\K_\v^{\pm}\rangle=\frac{1}{\sqrt{2}}(|d_{x^2-y^2}\rangle\pm \mathrm{i}|d_{xy}\rangle)$ orbit and the azimuthal component is $\langle \K_\v^{\pm}| \hat{L}_z |\K_\v^{\pm}\rangle=\pm 2\hbar$ \cite{https://www.doi.org/10.1103/PhysRevLett.108.196802}. Therefore the orbital term is $g(\mathrm{orbit})=(0 - (+2)) - (0 - (-2)) = -4$ . The valley term is proportional to the inverse of effective mass. For conduction band it is $g_{\c}^{V\pm,s}(\mathrm{valley})=\pm m_0/m_{\e}^{V}$, and for valence band it is $g_{v}^{V\pm,s}(\mathrm{valley})=\mp m_0/m_\h^{V}$, where $m_\e^{V}$ and $ m_\h^{V}$ are the effective masses of electron and hole at $V$ valley respectively. Thus the valley term is $g(\mathrm{valley})=(m_0/m_\e^{V} - m_0/m_\h^{V})-((-m_0/m_\e^{V}) - (-m_0/m_\h^{V})) = 2(m_0/m_\e^{V} - m_0/m_\h^{V})$\cite{https://doi.org/10.1038/s41467-017-01748-1}. In total, $g_{\X^0} = -4 + 2(m_0/m_\e^{\K} - m_0/m_\h^{\K})$. Because the effective masses of $m_\e$ and $m_\h$ are similar in WS${}_2$ .\cite{http://dx.doi.org/10.1063/1.4869142} Therefore, g factor of exciton is around $-4$, which has been confirmed experimentally before.\cite{http://dx.doi.org/10.1038/ncomms10643,http://dx.doi.org/10.1021/acs.nanolett.6b04171}

For negative trion, there are two electrons $\e_1, \e_2$ and one hole $\h$, where $\e_1$ at $V_1^\pm$ valley with spin $s_1^\pm$, $\e_2$ at $V_2^\pm$ valley with spin $s_2^\pm$, and $\h$ at $V_\h^\pm$ valley. Due to the large spin orbit coupling (SOC) in the valence band, excitons are clearly separated into A exciton and B exciton by energy difference. Here $\pm$ denotes hole spin direction of A exciton. If we neglect the influence of binding energy variation by magnetic field, only focus on Zeeman effect, then the g factor can be expressed as $g=g_\c^{V_j+,s_j+}-g_\v^{V_\h+,\uparrow}-g_\c^{V_j-,s_j-}+g_\v^{V_\h-,\downarrow}$, where $j=\{1,2\}$ is the electron index whose energy level is lower. Since there are two electrons, the excess electron of negative trion can falls in other valleys without violating the momentum conservation rule. If this excess electron possesses the same $\bm{k}$ vector but different spin, then the trion is singlet trion; if this extra electron falls in the opposite $\bm{k}$ vector but same spin, then the trion is triplet trion.
For singlet, $V_1^\pm=\K^{\pm}$, $V_2^\pm=\K^{\pm}$, $V_\h^\pm=\K^{\pm}$, $E(\e_1)>E(\e_2)$, $s_2^\pm=\mp\hbar/2$,
\begin{equation}
\begin{aligned}
g(\X^{-}_\mathrm{singlet}) &= g_\c^{\K+,\downarrow}-g_\v^{\K+,\uparrow}-g_\c^{\K-,\uparrow}+g_\v^{\K-,\downarrow}\\
&= ((-1/2) - (+1/2) - (+1/2) + (-1/2)) +\\
&+ (0 - (+2) - 0 + (-2)) +\\
&+ ((+\frac{m_0}{m_\e^\K}) - (+\frac{m_0}{m_\h^\K}) - (-\frac{m_0}{m_\e^\K}) + (-\frac{m_0}{m_\h^\K}))\\
&= -6 + 2(m_0/m_\e^\K - m_0/m_\h^\K) \approx -6
\end{aligned}
\end{equation}
For triplet, $V_1\pm=\K^{\pm}$, $V_2^\pm=\K^{\mp}$, $V_\h^\pm=\K^{\pm}$, $E(\e_1)>E(\e_2)$, $s_2^\pm=\pm\hbar/2$,
\begin{equation}
\begin{aligned}
g(\X^{-}_\mathrm{triplet}) &= g_\c^{\K-,\uparrow}-g_\v^{\K+,\uparrow}-g_\c^{\K+,\downarrow}+g_\v^{\K-,\downarrow}\\
&= ((+1/2) - (+1/2) - (-1/2) + (-1/2)) +\\
&+ (0 - (+2) - 0 + (-2)) +\\
&+ ((-\frac{m_0}{m_\e^\K}) - (+\frac{m_0}{m_\h^\K}) - (+\frac{m_0}{m_\e^\K}) + (-\frac{m_0}{m_\h^\K}))\\
&= -4 - 2(m_0/m_\e^\K + m_0/m_\h^\K) \ll -4.
\end{aligned}
\end{equation}
This is close to experimental values, for example in WSe${}_2$, the g factor is around $g=-5.3$ of singlet, and $-10.5$ of triplet.\cite{https://doi.org/10.1038/s41467-019-10228-7}

The $\Lambda$-K trions, whose extra electron falls in $\Lambda$ valley instead of K, have different orbital momentum from K-K trions, leading to different g factors. The electron state of conduction band is no more constituted of $|d_{z^2}\rangle$ orbit, but is a superposition of several states. To investigate the g factor of $\Lambda$-K trion, we estimated the orbital projections at $\Lambda$ valley, along with the band structure of monolayer WS${}_2$. The \textit{ab-initio} calculation is based on density functional theory. Projector augmented wave method \cite{https://link.aps.org/doi/10.1103/PhysRevB.50.17953} is used together with local density approximation. Four empty cell layers are filled between material layers to prevent interlayer interactions. The positions of atoms are initialized with structure data from Materials Project mp-224 .\cite{materialsproject} Before the electron density wave calculation, an ionic relaxation is executed to ensure the structure stable. The SOC is not considered in this calculation, since its influence to projection amplitudes is a minor term. The main influence of SOC is lifting the degeneracy of spins.

The result of projections on the orbital states around tungsten atoms are $1.6\%$ of $s$, $0.9\%$ of $p_x$, $0.7\%$ of $p_y$, $43.1\%$ of $d_{x^2-y^2}$, $32.5\%$ of $d_{xy}$, $21.1\%$ of $d_{z^2}$, and $0$ of else. For any state $|\psi\rangle$, if we have $\langle p_x|\psi\rangle=A_x \mathrm{exp}(\mathrm{i}\phi_x)$ and $\langle p_y|\psi\rangle=A_y \mathrm{exp}(\mathrm{i}\phi_y)$, then $|\langle p_+ | \psi\rangle|^2-|\langle p_- | \psi\rangle|^2 = 2 A_x A_y \mathrm{sin}(\phi_y - \phi_x)$, where $|p_{\pm}\rangle = \frac{1}{\sqrt{2}}(|p_x\rangle \pm\mathrm{i}|p_y\rangle)$. The phase difference will change to $\pm\frac{\pi}{2}$ when degeneracy is broken by magnet field or SOC. The amplitude of components thus are
\begin{equation}
\begin{aligned}
|\langle p_{+}|\psi_{\pm} \rangle|^2&-|\langle p_{-}|\psi_{\pm} \rangle|^2 = \pm 2|\langle p_{x}|\psi_{\pm} \rangle||\langle p_{y}|\psi_{\pm} \rangle|\\
 &= \pm 2\sqrt{0.9\%\times 0.7\%} = \pm 0.016\\
|\langle d_{(x+iy)^2}|\psi_{\pm} \rangle|^2 &- |\langle d_{(x-iy)^2}|\psi_{\pm} \rangle|^2=\\
&\pm  2|\langle d_{x^2-y^2}|\psi_{\pm} \rangle||\langle d_{xy}|\psi_{\pm} \rangle|\\
 &= \pm 2\sqrt{43.1\%\times 32.5\%} = \pm 0.7495\\
|\langle d_{(x\pm iy)z}|\psi_{\pm} \rangle|^2 &= 0.\\
\end{aligned}
\end{equation}
where $|\psi_{\pm}\rangle$ is the state of $\Lambda^{\pm}$ valley. Therefore the expected orbital angular momentum component in $z$ direction around tungsten atom should be:
\begin{equation}
\begin{aligned}
\langle \hat{L}_z \rangle &= 2\hbar|\langle d_{(x+iy)^2}|\psi \rangle|^2 - 2\hbar|\langle d_{(x-iy)^2}|\psi \rangle|^2 +\\
& + \hbar|\langle d_{(x+ iy)z} |\psi \rangle|^2 - \hbar|\langle d_{(x- iy)z}|\psi \rangle|^2 + 0 \hbar|\langle d_{z^2}|\psi \rangle|^2 +\\
& + \hbar|\langle p_{+} |\psi \rangle|^2 - \hbar|\langle p_{-}|\psi \rangle|^2 + 0 \hbar|\langle s |\psi \rangle|^2 \\
&= \pm\Big( 2 \times 0.7495 + 0 + 0 + 0.016 + 0 \Big) \hbar\\
&=\pm 1.515 \hbar.\\
\end{aligned}
\end{equation} And thus the conduction band energy shift of $\Lambda$ valley is larger than K valley
\begin{equation}
\label{eq:movement}
g_c^{\Lambda+,\uparrow} = \frac{1}{2} + 1.515 +\frac{m_0}{m_\e^\Lambda} > \frac{1}{2} + 0 +\frac{m_0}{m_\e^\K} = g_c^{\K+,\uparrow}.
\end{equation}
Therefore the energy difference between $\Lambda$ valley and K valley is reduced by magnetic field, which leads to the PL enhancement.

\begin{figure}[ht]
\centering
\includegraphics[scale=0.5]{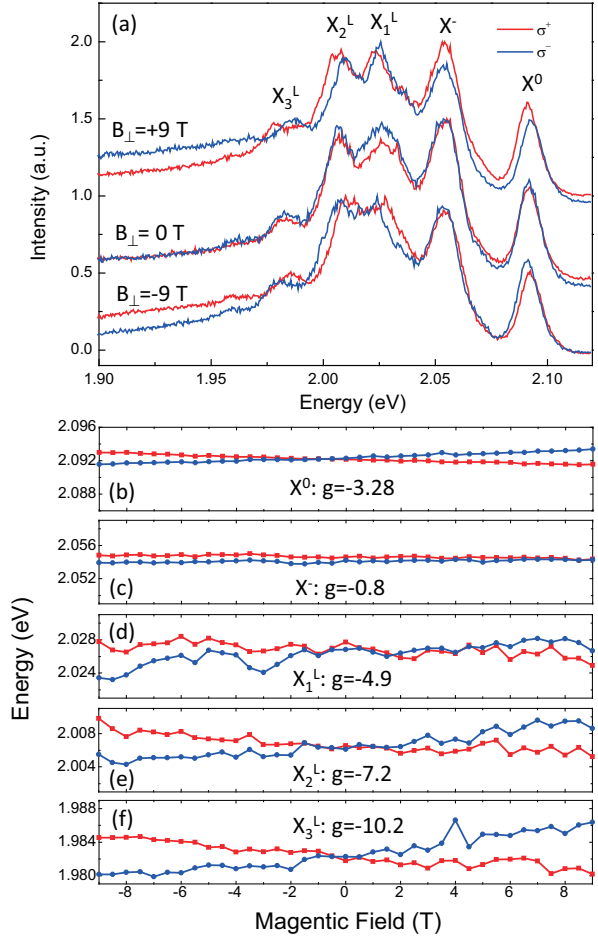}

\caption[]{\label{fig:gfactor}%
\raggedright
(a) The normalized spectra for 9 T, 0 T and -9 T. (b)-f) Peak centers of (b) X${}^0$, (c) X${}^-$ and (d-f) X${}_{1,2,3}^L$ for $\sigma^+$ (red) and $\sigma^-$ (blue) polarizations. The g factor of X${}^-$ is -0.8, whose absolute value is smaller than other excitons, while the g factors of X${}^0$ and X${}_{1,2,3}^L$ are consistent with previous reports.\cite{http://dx.doi.org/10.1038/ncomms10643,http://dx.doi.org/10.1021/acs.nanolett.6b04171,https://arxiv.xilesou.top/pdf/2002.11646.pdf}
}%
\end{figure}
As for g factor of $\Lambda$-K trion: $V_1^\pm=\K^{\pm}, V_2^\pm=\Lambda^{\pm}, V_\h^\pm=\K^{\pm}, E(\e_1)>E(\e_2), s_2^\pm=\pm\hbar/2, $
\begin{equation}
\begin{aligned}
g(\X^{-}_\mathrm{\Lambda-\K}) &= g_\c^{\Lambda+,\uparrow}-g_\v^{\K+,\uparrow}-g_\c^{\Lambda-,\downarrow}+g_\v^{\K-,\downarrow}\\
&= ((+1/2) - (+1/2) - (-1/2) + (-1/2)) +\\
&+ ((+1.515) - (+2) - (-1.515) + (-2)) +\\
&+ ((+\frac{m_0}{m_\e^\Lambda}) - (+\frac{m_0}{m_\h^\K}) - (-\frac{m_0}{m_\e^\Lambda}) + (-\frac{m_0}{m_\h^\K}))\\
&= -0.970 - 2(m_0/m_\e^\Lambda - m_0/m_\h^\K).
\end{aligned}
\end{equation}
Since the effective masses of electron and hole are similar, we neglect the $g_{\mathrm{valley}}$ in the same way as K-K trion. Therefore $g\approx -0.97$, whose absolute value is extraordinarily small.

In order to obtain Land\'{e} g factor of the exciton states experimentally, we measured $\sigma^{+}$ and $\sigma^{-}$ components of the PL spectra at different magnetic fields. The normalized spectra for $9$ T, $0$ T and $-9$ T are shown in Fig. \ref{fig:gfactor} (a). The red curves are $\sigma^{+}$ polarization and the blue curves are $\sigma^{-}$ polarization. The peak centers are extracted by multi-peak non-linear least-squares curve-fitting python package from spectra. And the g factor is extracted by linear regression of the peak centers. The measured g factor of X${}^0$ is $-3.08$, close to theoretical value (around $-4$).\cite{http://dx.doi.org/10.1038/ncomms10643,http://dx.doi.org/10.1021/acs.nanolett.6b04171} The measured g factors of X${}_{1,2,3}^L$ are $-4.9$, $-7.2$ and $-10.2$ respectively, consists with reported values in TMDs.\cite{https://arxiv.xilesou.top/pdf/2002.11646.pdf} So far, the g factors of defect-bound excitons vary in a wide range from $-6$ to $-16$ which have been reported in TMDs.\cite{https://aip.scitation.org/doi/10.1063/1.5087440,https://doi.org/10.1038/nnano.2015.75,https://doi.org/10.1038/nnano.2015.60} Our results for defect related peaks fall in this range. Surprisingly, the measured g factor of X${}^-$ is $-0.8$ whose absolute value is much smaller than above values as shown in Fig. \ref{fig:gfactor} (c). The small absolute value is close to our calculation $|-0.97|$, which confirms our assumption that the trion is an indirect-transition between $\Lambda$ and K valleys. And with the assistence of $\Lambda$-K trions, the PL intensity can be enhanced.

\section{\label{sec:conclusion}Conclusions}
In conclusion, a magnetic field induced large PL enhancement of WS${}_2$ in cryogenic environment is reported. Two mechanisms have been discussed to explain the enhancement qualitatively. One is attributed to the magnetic field induced wavefunction confinement causing a larger overlap of electron and hole wavefunction extension. And the other is related to an indirect-transition trion between $\Lambda$ and K valleys. According to the model considered, the $\Lambda$-K trion has a smaller absolute value of g factor than K-K trions, which is confirmed experimentally. We believe this work will extend valleytronics with different valleys in monolayer TMDs for future.

\begin{acknowledgments}
This work was supported by the National Natural Science Foundation of China under Grants Nos. 11934019, 61675228, 11721404, 51761145104 and 11874419; the Strategic Priority Research Program, the Instrument Developing Project and the Interdisciplinary Innovation Team of the Chinese Academy of Sciences under Grants Nos. XDB28000000 and YJKYYQ20180036, and the Key Research and Development Program of Guangdong Province under Grant No. 2018B030329001.
\end{acknowledgments}


%







\begin{thebibliography}{46}%
\makeatletter
\providecommand \@ifxundefined [1]{%
 \@ifx{#1\undefined}
}%
\providecommand \@ifnum [1]{%
 \ifnum #1\expandafter \@firstoftwo
 \else \expandafter \@secondoftwo
 \fi
}%
\providecommand \@ifx [1]{%
 \ifx #1\expandafter \@firstoftwo
 \else \expandafter \@secondoftwo
 \fi
}%
\providecommand \natexlab [1]{#1}%
\providecommand \enquote  [1]{``#1''}%
\providecommand \bibnamefont  [1]{#1}%
\providecommand \bibfnamefont [1]{#1}%
\providecommand \citenamefont [1]{#1}%
\providecommand \href@noop [0]{\@secondoftwo}%
\providecommand \href [0]{\begingroup \@sanitize@url \@href}%
\providecommand \@href[1]{\@@startlink{#1}\@@href}%
\providecommand \@@href[1]{\endgroup#1\@@endlink}%
\providecommand \@sanitize@url [0]{\catcode `\\12\catcode `\$12\catcode
  `\&12\catcode `\#12\catcode `\^12\catcode `\_12\catcode `\%12\relax}%
\providecommand \@@startlink[1]{}%
\providecommand \@@endlink[0]{}%
\providecommand \url  [0]{\begingroup\@sanitize@url \@url }%
\providecommand \@url [1]{\endgroup\@href {#1}{\urlprefix }}%
\providecommand \urlprefix  [0]{URL }%
\providecommand \Eprint [0]{\href }%
\providecommand \doibase [0]{http://dx.doi.org/}%
\providecommand \selectlanguage [0]{\@gobble}%
\providecommand \bibinfo  [0]{\@secondoftwo}%
\providecommand \bibfield  [0]{\@secondoftwo}%
\providecommand \translation [1]{[#1]}%
\providecommand \BibitemOpen [0]{}%
\providecommand \bibitemStop [0]{}%
\providecommand \bibitemNoStop [0]{.\EOS\space}%
\providecommand \EOS [0]{\spacefactor3000\relax}%
\providecommand \BibitemShut  [1]{\csname bibitem#1\endcsname}%
\let\auto@bib@innerbib\@empty
\bibitem [{\citenamefont {Xiao}\ \emph {et~al.}(2012)\citenamefont {Xiao},
  \citenamefont {Liu}, \citenamefont {Feng}, \citenamefont {Xu},\ and\
  \citenamefont {Yao}}]{https://www.doi.org/10.1103/PhysRevLett.108.196802}%
  \BibitemOpen
  \bibfield  {author} {\bibinfo {author} {\bibfnamefont {D.}~\bibnamefont
  {Xiao}}, \bibinfo {author} {\bibfnamefont {G.~B.}\ \bibnamefont {Liu}},
  \bibinfo {author} {\bibfnamefont {W.}~\bibnamefont {Feng}}, \bibinfo {author}
  {\bibfnamefont {X.}~\bibnamefont {Xu}}, \ and\ \bibinfo {author}
  {\bibfnamefont {W.}~\bibnamefont {Yao}},\ }\href {\doibase
  10.1103/PhysRevLett.108.196802} {\bibfield  {journal} {\bibinfo  {journal}
  {Phys. Rev. Lett.}\ }\textbf {\bibinfo {volume} {108}},\ \bibinfo {pages}
  {196802} (\bibinfo {year} {2012})}\BibitemShut {NoStop}%
\bibitem [{\citenamefont {Mak}\ \emph {et~al.}(2010)\citenamefont {Mak},
  \citenamefont {Lee}, \citenamefont {Hone}, \citenamefont {Shan},\ and\
  \citenamefont {Heinz}}]{PhysRevLett.105.136805}%
  \BibitemOpen
  \bibfield  {author} {\bibinfo {author} {\bibfnamefont {K.~F.}\ \bibnamefont
  {Mak}}, \bibinfo {author} {\bibfnamefont {C.}~\bibnamefont {Lee}}, \bibinfo
  {author} {\bibfnamefont {J.}~\bibnamefont {Hone}}, \bibinfo {author}
  {\bibfnamefont {J.}~\bibnamefont {Shan}}, \ and\ \bibinfo {author}
  {\bibfnamefont {T.~F.}\ \bibnamefont {Heinz}},\ }\href {\doibase
  10.1103/PhysRevLett.105.136805} {\bibfield  {journal} {\bibinfo  {journal}
  {Phys. Rev. Lett.}\ }\textbf {\bibinfo {volume} {105}},\ \bibinfo {pages}
  {136805} (\bibinfo {year} {2010})}\BibitemShut {NoStop}%
\bibitem [{\citenamefont {Splendiani}\ \emph {et~al.}(2010)\citenamefont
  {Splendiani}, \citenamefont {Sun}, \citenamefont {Zhang}, \citenamefont {Li},
  \citenamefont {Kim}, \citenamefont {Chim}, \citenamefont {Galli},\ and\
  \citenamefont {Wang}}]{doi:10.1021/nl903868w}%
  \BibitemOpen
  \bibfield  {author} {\bibinfo {author} {\bibfnamefont {A.}~\bibnamefont
  {Splendiani}}, \bibinfo {author} {\bibfnamefont {L.}~\bibnamefont {Sun}},
  \bibinfo {author} {\bibfnamefont {Y.}~\bibnamefont {Zhang}}, \bibinfo
  {author} {\bibfnamefont {T.}~\bibnamefont {Li}}, \bibinfo {author}
  {\bibfnamefont {J.}~\bibnamefont {Kim}}, \bibinfo {author} {\bibfnamefont
  {C.-Y.}\ \bibnamefont {Chim}}, \bibinfo {author} {\bibfnamefont
  {G.}~\bibnamefont {Galli}}, \ and\ \bibinfo {author} {\bibfnamefont
  {F.}~\bibnamefont {Wang}},\ }\href@noop {} {\bibfield  {journal} {\bibinfo
  {journal} {Nano Lett.}\ }\textbf {\bibinfo {volume} {10}},\ \bibinfo {pages}
  {1271} (\bibinfo {year} {2010})}\BibitemShut {NoStop}%
\bibitem [{\citenamefont {Zhao}\ \emph {et~al.}(2013)\citenamefont {Zhao},
  \citenamefont {Ghorannevis}, \citenamefont {Chu}, \citenamefont {Toh},
  \citenamefont {Kloc}, \citenamefont {Tan},\ and\ \citenamefont
  {Eda}}]{doi:10.1021/nn305275h}%
  \BibitemOpen
  \bibfield  {author} {\bibinfo {author} {\bibfnamefont {W.}~\bibnamefont
  {Zhao}}, \bibinfo {author} {\bibfnamefont {Z.}~\bibnamefont {Ghorannevis}},
  \bibinfo {author} {\bibfnamefont {L.}~\bibnamefont {Chu}}, \bibinfo {author}
  {\bibfnamefont {M.}~\bibnamefont {Toh}}, \bibinfo {author} {\bibfnamefont
  {C.}~\bibnamefont {Kloc}}, \bibinfo {author} {\bibfnamefont {P.-H.}\
  \bibnamefont {Tan}}, \ and\ \bibinfo {author} {\bibfnamefont
  {G.}~\bibnamefont {Eda}},\ }\href@noop {} {\bibfield  {journal} {\bibinfo
  {journal} {ACS Nano.}\ }\textbf {\bibinfo {volume} {7}},\ \bibinfo {pages}
  {791} (\bibinfo {year} {2013})}\BibitemShut {NoStop}%
\bibitem [{\citenamefont {Allain}\ and\ \citenamefont
  {Kis}(2014)}]{doi:10.1021/nn5021538}%
  \BibitemOpen
  \bibfield  {author} {\bibinfo {author} {\bibfnamefont {A.}~\bibnamefont
  {Allain}}\ and\ \bibinfo {author} {\bibfnamefont {A.}~\bibnamefont {Kis}},\
  }\href@noop {} {\bibfield  {journal} {\bibinfo  {journal} {ACS Nano.}\
  }\textbf {\bibinfo {volume} {8}},\ \bibinfo {pages} {7180} (\bibinfo {year}
  {2014})}\BibitemShut {NoStop}%
\bibitem [{\citenamefont {Srivastava}\ \emph
  {et~al.}(2015{\natexlab{a}})\citenamefont {Srivastava}, \citenamefont
  {Sidler}, \citenamefont {Allain}, \citenamefont {Lembke}, \citenamefont
  {Kis},\ and\ \citenamefont {Imamoglu}}]{ISI:000349934700017}%
  \BibitemOpen
  \bibfield  {author} {\bibinfo {author} {\bibfnamefont {A.}~\bibnamefont
  {Srivastava}}, \bibinfo {author} {\bibfnamefont {M.}~\bibnamefont {Sidler}},
  \bibinfo {author} {\bibfnamefont {A.~V.}\ \bibnamefont {Allain}}, \bibinfo
  {author} {\bibfnamefont {D.~S.}\ \bibnamefont {Lembke}}, \bibinfo {author}
  {\bibfnamefont {A.}~\bibnamefont {Kis}}, \ and\ \bibinfo {author}
  {\bibfnamefont {A.}~\bibnamefont {Imamoglu}},\ }\href@noop {} {\bibfield
  {journal} {\bibinfo  {journal} {Nat. Phys.}\ }\textbf {\bibinfo {volume}
  {11}},\ \bibinfo {pages} {141} (\bibinfo {year}
  {2015}{\natexlab{a}})}\BibitemShut {NoStop}%
\bibitem [{\citenamefont {Wang}\ \emph
  {et~al.}(2015{\natexlab{a}})\citenamefont {Wang}, \citenamefont {Bouet},
  \citenamefont {Glazov}, \citenamefont {Amand}, \citenamefont {Ivchenko},
  \citenamefont {Palleau}, \citenamefont {Marie},\ and\ \citenamefont
  {Urbaszek}}]{ISI:000367748600015}%
  \BibitemOpen
  \bibfield  {author} {\bibinfo {author} {\bibfnamefont {G.}~\bibnamefont
  {Wang}}, \bibinfo {author} {\bibfnamefont {L.}~\bibnamefont {Bouet}},
  \bibinfo {author} {\bibfnamefont {M.~M.}\ \bibnamefont {Glazov}}, \bibinfo
  {author} {\bibfnamefont {T.}~\bibnamefont {Amand}}, \bibinfo {author}
  {\bibfnamefont {E.~L.}\ \bibnamefont {Ivchenko}}, \bibinfo {author}
  {\bibfnamefont {E.}~\bibnamefont {Palleau}}, \bibinfo {author} {\bibfnamefont
  {X.}~\bibnamefont {Marie}}, \ and\ \bibinfo {author} {\bibfnamefont
  {B.}~\bibnamefont {Urbaszek}},\ }\href@noop {} {\bibfield  {journal}
  {\bibinfo  {journal} {{2D Mater.}}\ }\textbf {\bibinfo {volume} {{2}}},\
  \bibinfo {pages} {{3}} (\bibinfo {year} {{2015}}{\natexlab{a}})}\BibitemShut
  {NoStop}%
\bibitem [{\citenamefont {F{\"o}rste}\ \emph {et~al.}(2020)\citenamefont
  {F{\"o}rste}, \citenamefont {Tepliakov}, \citenamefont {Kruchinin},
  \citenamefont {Lindlau}, \citenamefont {Funk}, \citenamefont {F{\"o}rg},
  \citenamefont {Watanabe}, \citenamefont {Taniguchi}, \citenamefont
  {Baimuratov},\ and\ \citenamefont
  {H{\"o}gele}}]{https://arxiv.xilesou.top/pdf/2002.11646.pdf}%
  \BibitemOpen
  \bibfield  {author} {\bibinfo {author} {\bibfnamefont {J.}~\bibnamefont
  {F{\"o}rste}}, \bibinfo {author} {\bibfnamefont {N.~V.}\ \bibnamefont
  {Tepliakov}}, \bibinfo {author} {\bibfnamefont {S.~Y.}\ \bibnamefont
  {Kruchinin}}, \bibinfo {author} {\bibfnamefont {J.}~\bibnamefont {Lindlau}},
  \bibinfo {author} {\bibfnamefont {V.}~\bibnamefont {Funk}}, \bibinfo {author}
  {\bibfnamefont {M.}~\bibnamefont {F{\"o}rg}}, \bibinfo {author}
  {\bibfnamefont {K.}~\bibnamefont {Watanabe}}, \bibinfo {author}
  {\bibfnamefont {T.}~\bibnamefont {Taniguchi}}, \bibinfo {author}
  {\bibfnamefont {A.~S.}\ \bibnamefont {Baimuratov}}, \ and\ \bibinfo {author}
  {\bibfnamefont {A.}~\bibnamefont {H{\"o}gele}},\ }\href
  {https://arxiv.xilesou.top/pdf/2002.11646} {\bibfield  {journal} {\bibinfo
  {journal} {arXiv print arXiv:2002.11646}\ } (\bibinfo {year}
  {2020})}\BibitemShut {NoStop}%
\bibitem [{\citenamefont {Rybkovskiy}, \citenamefont {Gerber},\ and\
  \citenamefont {Durnev}(2017)}]{ISI:000412635500004}%
  \BibitemOpen
  \bibfield  {author} {\bibinfo {author} {\bibfnamefont {D.~V.}\ \bibnamefont
  {Rybkovskiy}}, \bibinfo {author} {\bibfnamefont {I.~C.}\ \bibnamefont
  {Gerber}}, \ and\ \bibinfo {author} {\bibfnamefont {M.~V.}\ \bibnamefont
  {Durnev}},\ }\href@noop {} {\bibfield  {journal} {\bibinfo  {journal} {{Phys.
  Rev. B}}\ }\textbf {\bibinfo {volume} {{95}}},\ \bibinfo {pages} {{15}}
  (\bibinfo {year} {{2017}})}\BibitemShut {NoStop}%
\bibitem [{\citenamefont {Wu}\ \emph {et~al.}(2018)\citenamefont {Wu},
  \citenamefont {Shen}, \citenamefont {Tan}, \citenamefont {Zhang},
  \citenamefont {Tan},\ and\ \citenamefont {Zheng}}]{10.7498/aps.67.20180615}%
  \BibitemOpen
  \bibfield  {author} {\bibinfo {author} {\bibfnamefont {Y.-J.}\ \bibnamefont
  {Wu}}, \bibinfo {author} {\bibfnamefont {C.}~\bibnamefont {Shen}}, \bibinfo
  {author} {\bibfnamefont {Q.-H.}\ \bibnamefont {Tan}}, \bibinfo {author}
  {\bibfnamefont {J.}~\bibnamefont {Zhang}}, \bibinfo {author} {\bibfnamefont
  {P.-H.}\ \bibnamefont {Tan}}, \ and\ \bibinfo {author} {\bibfnamefont
  {H.-Z.}\ \bibnamefont {Zheng}},\ }\href {\doibase 10.7498/aps.67.20180615}
  {\bibfield  {journal} {\bibinfo  {journal} {Acta Physica Sinica}\ }\textbf
  {\bibinfo {volume} {67}},\ \bibinfo {pages} {147801} (\bibinfo {year}
  {2018})}\BibitemShut {NoStop}%
\bibitem [{\citenamefont {Koperski}\ \emph {et~al.}(2017)\citenamefont
  {Koperski}, \citenamefont {Molas}, \citenamefont {Arora}, \citenamefont
  {Nogajewski}, \citenamefont {Slobodeniuk}, \citenamefont {Faugeras},\ and\
  \citenamefont {Potemski}}]{https://doi.org/10.1515/nanoph-2016-0165}%
  \BibitemOpen
  \bibfield  {author} {\bibinfo {author} {\bibfnamefont {M.}~\bibnamefont
  {Koperski}}, \bibinfo {author} {\bibfnamefont {M.~R.}\ \bibnamefont {Molas}},
  \bibinfo {author} {\bibfnamefont {A.}~\bibnamefont {Arora}}, \bibinfo
  {author} {\bibfnamefont {K.}~\bibnamefont {Nogajewski}}, \bibinfo {author}
  {\bibfnamefont {A.~O.}\ \bibnamefont {Slobodeniuk}}, \bibinfo {author}
  {\bibfnamefont {C.}~\bibnamefont {Faugeras}}, \ and\ \bibinfo {author}
  {\bibfnamefont {M.}~\bibnamefont {Potemski}},\ }\href
  {https://doi.org/10.1515/nanoph-2016-0165} {\bibfield  {journal} {\bibinfo
  {journal} {Nanophotonics}\ }\textbf {\bibinfo {volume} {6}},\ \bibinfo
  {pages} {1289} (\bibinfo {year} {2017})}\BibitemShut {NoStop}%
\bibitem [{\citenamefont {Liu}\ \emph {et~al.}(2013)\citenamefont {Liu},
  \citenamefont {Shan}, \citenamefont {Yao}, \citenamefont {Yao},\ and\
  \citenamefont {Xiao}}]{https://www.doi.org/10.1103/PhysRevB.88.085433}%
  \BibitemOpen
  \bibfield  {author} {\bibinfo {author} {\bibfnamefont {G.~B.}\ \bibnamefont
  {Liu}}, \bibinfo {author} {\bibfnamefont {W.~Y.}\ \bibnamefont {Shan}},
  \bibinfo {author} {\bibfnamefont {Y.}~\bibnamefont {Yao}}, \bibinfo {author}
  {\bibfnamefont {W.}~\bibnamefont {Yao}}, \ and\ \bibinfo {author}
  {\bibfnamefont {D.}~\bibnamefont {Xiao}},\ }\href {\doibase
  10.1103/PhysRevB.88.085433} {\bibfield  {journal} {\bibinfo  {journal} {Phys.
  Rev. B}\ }\textbf {\bibinfo {volume} {88}},\ \bibinfo {pages} {085433}
  (\bibinfo {year} {2013})}\BibitemShut {NoStop}%
\bibitem [{\citenamefont {Molas}\ \emph {et~al.}(2017)\citenamefont {Molas},
  \citenamefont {Faugeras}, \citenamefont {Slobodeniuk}, \citenamefont
  {Nogajewski}, \citenamefont {Bartos}, \citenamefont {Basko},\ and\
  \citenamefont {Potemski}}]{https://www.doi.org/10.1088/2053-1583/aa5521}%
  \BibitemOpen
  \bibfield  {author} {\bibinfo {author} {\bibfnamefont {M.~R.}\ \bibnamefont
  {Molas}}, \bibinfo {author} {\bibfnamefont {C.}~\bibnamefont {Faugeras}},
  \bibinfo {author} {\bibfnamefont {A.~O.}\ \bibnamefont {Slobodeniuk}},
  \bibinfo {author} {\bibfnamefont {K.}~\bibnamefont {Nogajewski}}, \bibinfo
  {author} {\bibfnamefont {M.}~\bibnamefont {Bartos}}, \bibinfo {author}
  {\bibfnamefont {D.~M.}\ \bibnamefont {Basko}}, \ and\ \bibinfo {author}
  {\bibfnamefont {M.}~\bibnamefont {Potemski}},\ }\href {\doibase
  10.1088/2053-1583/aa5521} {\bibfield  {journal} {\bibinfo  {journal} {2D
  Mater.}\ }\textbf {\bibinfo {volume} {4}},\ \bibinfo {pages} {021003}
  (\bibinfo {year} {2017})}\BibitemShut {NoStop}%
\bibitem [{\citenamefont {Zhang}\ \emph {et~al.}(2017)\citenamefont {Zhang},
  \citenamefont {Cao}, \citenamefont {Lu}, \citenamefont {Lin}, \citenamefont
  {Zhang}, \citenamefont {Wang}, \citenamefont {Li}, \citenamefont {Hone},
  \citenamefont {Robinson}, \citenamefont {Smirnov}, \citenamefont {Louie},\
  and\ \citenamefont {Heinz}}]{https://www.doi.org/10.1038/NNANO.2017.105}%
  \BibitemOpen
  \bibfield  {author} {\bibinfo {author} {\bibfnamefont {X.~X.}\ \bibnamefont
  {Zhang}}, \bibinfo {author} {\bibfnamefont {T.}~\bibnamefont {Cao}}, \bibinfo
  {author} {\bibfnamefont {Z.~G.}\ \bibnamefont {Lu}}, \bibinfo {author}
  {\bibfnamefont {Y.~C.}\ \bibnamefont {Lin}}, \bibinfo {author} {\bibfnamefont
  {F.}~\bibnamefont {Zhang}}, \bibinfo {author} {\bibfnamefont
  {Y.}~\bibnamefont {Wang}}, \bibinfo {author} {\bibfnamefont {Z.}~\bibnamefont
  {Li}}, \bibinfo {author} {\bibfnamefont {J.~C.}\ \bibnamefont {Hone}},
  \bibinfo {author} {\bibfnamefont {J.~A.}\ \bibnamefont {Robinson}}, \bibinfo
  {author} {\bibfnamefont {D.}~\bibnamefont {Smirnov}}, \bibinfo {author}
  {\bibfnamefont {S.~G.}\ \bibnamefont {Louie}}, \ and\ \bibinfo {author}
  {\bibfnamefont {T.~F.}\ \bibnamefont {Heinz}},\ }\href
  {https://www.doi.org/10.1038/NNANO.2017.105} {\bibfield  {journal} {\bibinfo
  {journal} {Nat. Nanotech.}\ }\textbf {\bibinfo {volume} {12}},\ \bibinfo
  {pages} {883} (\bibinfo {year} {2017})}\BibitemShut {NoStop}%
\bibitem [{Note1()}]{Note1}%
  \BibitemOpen
  \bibinfo {note} {Q and $\Lambda $ points in the Brillouin zone have common
  $x,y$ but different $z$. For monolayer 2D materials, Q and $\Lambda $ can be
  treated as the same.}\BibitemShut {Stop}%
\bibitem [{\citenamefont {Lindlau}\ \emph {et~al.}(2017)\citenamefont
  {Lindlau}, \citenamefont {Robert}, \citenamefont {Funk}, \citenamefont
  {Förste}, \citenamefont {Förg}, \citenamefont {Colombier}, \citenamefont
  {Neumann}, \citenamefont {Courtade}, \citenamefont {Shree}, \citenamefont
  {Taniguchi}, \citenamefont {Watanabe}, \citenamefont {Glazov}, \citenamefont
  {Marie}, \citenamefont {Urbaszek},\ and\ \citenamefont
  {Högele}}]{lindlau2017identifying}%
  \BibitemOpen
  \bibfield  {author} {\bibinfo {author} {\bibfnamefont {J.}~\bibnamefont
  {Lindlau}}, \bibinfo {author} {\bibfnamefont {C.}~\bibnamefont {Robert}},
  \bibinfo {author} {\bibfnamefont {V.}~\bibnamefont {Funk}}, \bibinfo {author}
  {\bibfnamefont {J.}~\bibnamefont {Förste}}, \bibinfo {author} {\bibfnamefont
  {M.}~\bibnamefont {Förg}}, \bibinfo {author} {\bibfnamefont
  {L.}~\bibnamefont {Colombier}}, \bibinfo {author} {\bibfnamefont
  {A.}~\bibnamefont {Neumann}}, \bibinfo {author} {\bibfnamefont
  {E.}~\bibnamefont {Courtade}}, \bibinfo {author} {\bibfnamefont
  {S.}~\bibnamefont {Shree}}, \bibinfo {author} {\bibfnamefont
  {T.}~\bibnamefont {Taniguchi}}, \bibinfo {author} {\bibfnamefont
  {K.}~\bibnamefont {Watanabe}}, \bibinfo {author} {\bibfnamefont {M.~M.}\
  \bibnamefont {Glazov}}, \bibinfo {author} {\bibfnamefont {X.}~\bibnamefont
  {Marie}}, \bibinfo {author} {\bibfnamefont {B.}~\bibnamefont {Urbaszek}}, \
  and\ \bibinfo {author} {\bibfnamefont {A.}~\bibnamefont {Högele}},\
  }\href@noop {} {\bibfield  {journal} {\bibinfo  {journal} {arXiv print
  arXiv:1710.00988}\ } (\bibinfo {year} {2017})}\BibitemShut {NoStop}%
\bibitem [{\citenamefont {Desai}\ \emph {et~al.}(2014)\citenamefont {Desai},
  \citenamefont {Seol}, \citenamefont {Kang}, \citenamefont {Fang},
  \citenamefont {Battaglia}, \citenamefont {Kapadia}, \citenamefont {Ager},
  \citenamefont {Guo},\ and\ \citenamefont
  {Javey}}]{https://doi.org/10.1021/nl501638a}%
  \BibitemOpen
  \bibfield  {author} {\bibinfo {author} {\bibfnamefont {S.~B.}\ \bibnamefont
  {Desai}}, \bibinfo {author} {\bibfnamefont {G.}~\bibnamefont {Seol}},
  \bibinfo {author} {\bibfnamefont {J.~S.}\ \bibnamefont {Kang}}, \bibinfo
  {author} {\bibfnamefont {H.}~\bibnamefont {Fang}}, \bibinfo {author}
  {\bibfnamefont {C.}~\bibnamefont {Battaglia}}, \bibinfo {author}
  {\bibfnamefont {R.}~\bibnamefont {Kapadia}}, \bibinfo {author} {\bibfnamefont
  {J.~W.}\ \bibnamefont {Ager}}, \bibinfo {author} {\bibfnamefont
  {J.}~\bibnamefont {Guo}}, \ and\ \bibinfo {author} {\bibfnamefont
  {A.}~\bibnamefont {Javey}},\ }\href {\doibase 10.1021/nl501638a} {\bibfield
  {journal} {\bibinfo  {journal} {Nano Lett.}\ }\textbf {\bibinfo {volume}
  {14}},\ \bibinfo {pages} {4592} (\bibinfo {year} {2014})}\BibitemShut
  {NoStop}%
\bibitem [{\citenamefont {Wang}\ \emph
  {et~al.}(2015{\natexlab{b}})\citenamefont {Wang}, \citenamefont {Cong},
  \citenamefont {Yang}, \citenamefont {Shang}, \citenamefont {Peimyoo},
  \citenamefont {Chen}, \citenamefont {Kang}, \citenamefont {Wang},
  \citenamefont {Huang},\ and\ \citenamefont
  {Yu}}]{https://doi.org/10.1007/s12274-015-0762-6}%
  \BibitemOpen
  \bibfield  {author} {\bibinfo {author} {\bibfnamefont {Y.}~\bibnamefont
  {Wang}}, \bibinfo {author} {\bibfnamefont {C.}~\bibnamefont {Cong}}, \bibinfo
  {author} {\bibfnamefont {W.}~\bibnamefont {Yang}}, \bibinfo {author}
  {\bibfnamefont {J.}~\bibnamefont {Shang}}, \bibinfo {author} {\bibfnamefont
  {N.}~\bibnamefont {Peimyoo}}, \bibinfo {author} {\bibfnamefont
  {Y.}~\bibnamefont {Chen}}, \bibinfo {author} {\bibfnamefont {J.}~\bibnamefont
  {Kang}}, \bibinfo {author} {\bibfnamefont {J.}~\bibnamefont {Wang}}, \bibinfo
  {author} {\bibfnamefont {W.}~\bibnamefont {Huang}}, \ and\ \bibinfo {author}
  {\bibfnamefont {T.}~\bibnamefont {Yu}},\ }\href {\doibase
  10.1007/s12274-015-0762-6} {\bibfield  {journal} {\bibinfo  {journal} {Nano
  Res.}\ }\textbf {\bibinfo {volume} {8}},\ \bibinfo {pages} {2562} (\bibinfo
  {year} {2015}{\natexlab{b}})}\BibitemShut {NoStop}%
\bibitem [{\citenamefont {Wickramaratne}, \citenamefont {Zahid},\ and\
  \citenamefont {Lake}(2014)}]{http://dx.doi.org/10.1063/1.4869142}%
  \BibitemOpen
  \bibfield  {author} {\bibinfo {author} {\bibfnamefont {D.}~\bibnamefont
  {Wickramaratne}}, \bibinfo {author} {\bibfnamefont {F.}~\bibnamefont
  {Zahid}}, \ and\ \bibinfo {author} {\bibfnamefont {R.~K.}\ \bibnamefont
  {Lake}},\ }\href {http://dx.doi.org/10.1063/1.4869142} {\bibfield  {journal}
  {\bibinfo  {journal} {J. Chem. Phys.}\ }\textbf {\bibinfo {volume} {140}},\
  \bibinfo {pages} {124710} (\bibinfo {year} {2014})}\BibitemShut {NoStop}%
\bibitem [{\citenamefont {{Plechinger}}\ \emph {et~al.}(2015)\citenamefont
  {{Plechinger}}, \citenamefont {{Nagler}}, \citenamefont {{Kraus}},
  \citenamefont {{Paradiso}}, \citenamefont {{Strunk}}, \citenamefont
  {{Sch{\"u}ller}},\ and\ \citenamefont
  {{Korn}}}]{https://doi.org/10.1002/pssr.201510224}%
  \BibitemOpen
  \bibfield  {author} {\bibinfo {author} {\bibfnamefont {G.}~\bibnamefont
  {{Plechinger}}}, \bibinfo {author} {\bibfnamefont {P.}~\bibnamefont
  {{Nagler}}}, \bibinfo {author} {\bibfnamefont {J.}~\bibnamefont {{Kraus}}},
  \bibinfo {author} {\bibfnamefont {N.}~\bibnamefont {{Paradiso}}}, \bibinfo
  {author} {\bibfnamefont {C.}~\bibnamefont {{Strunk}}}, \bibinfo {author}
  {\bibfnamefont {C.}~\bibnamefont {{Sch{\"u}ller}}}, \ and\ \bibinfo {author}
  {\bibfnamefont {T.}~\bibnamefont {{Korn}}},\ }\href
  {https://doi.org/10.1002/pssr.201510224} {\bibfield  {journal} {\bibinfo
  {journal} {Phys. Status Solidi RRL}\ }\textbf {\bibinfo {volume} {9}},\
  \bibinfo {pages} {457} (\bibinfo {year} {2015})}\BibitemShut {NoStop}%
\bibitem [{\citenamefont {Chernikov}\ \emph {et~al.}(2015)\citenamefont
  {Chernikov}, \citenamefont {van~der Zande}, \citenamefont {Hill},
  \citenamefont {Rigosi}, \citenamefont {Velauthapillai}, \citenamefont
  {Hone},\ and\ \citenamefont
  {Heinz}}]{http://dx.doi.org/10.1103/PhysRevLett.115.126802}%
  \BibitemOpen
  \bibfield  {author} {\bibinfo {author} {\bibfnamefont {A.}~\bibnamefont
  {Chernikov}}, \bibinfo {author} {\bibfnamefont {A.~M.}\ \bibnamefont {van~der
  Zande}}, \bibinfo {author} {\bibfnamefont {H.~M.}\ \bibnamefont {Hill}},
  \bibinfo {author} {\bibfnamefont {A.~F.}\ \bibnamefont {Rigosi}}, \bibinfo
  {author} {\bibfnamefont {A.}~\bibnamefont {Velauthapillai}}, \bibinfo
  {author} {\bibfnamefont {J.}~\bibnamefont {Hone}}, \ and\ \bibinfo {author}
  {\bibfnamefont {T.~F.}\ \bibnamefont {Heinz}},\ }\href {\doibase
  10.1103/PhysRevLett.115.126802} {\bibfield  {journal} {\bibinfo  {journal}
  {Phys. Rev. Lett.}\ }\textbf {\bibinfo {volume} {115}},\ \bibinfo {pages}
  {126802} (\bibinfo {year} {2015})}\BibitemShut {NoStop}%
\bibitem [{\citenamefont {Zhu}, \citenamefont {Chen},\ and\ \citenamefont
  {Cui}(2015)}]{https://doi.org/10.1038/srep09218}%
  \BibitemOpen
  \bibfield  {author} {\bibinfo {author} {\bibfnamefont {B.}~\bibnamefont
  {Zhu}}, \bibinfo {author} {\bibfnamefont {X.}~\bibnamefont {Chen}}, \ and\
  \bibinfo {author} {\bibfnamefont {X.}~\bibnamefont {Cui}},\ }\href {\doibase
  10.1038/srep09218} {\bibfield  {journal} {\bibinfo  {journal} {Sci. Rep.}\
  }\textbf {\bibinfo {volume} {5}},\ \bibinfo {pages} {9218} (\bibinfo {year}
  {2015})}\BibitemShut {NoStop}%
\bibitem [{\citenamefont {Barbone}\ \emph {et~al.}(2018)\citenamefont
  {Barbone}, \citenamefont {Montblanch}, \citenamefont {Kara}, \citenamefont
  {Palacios-Berraquero}, \citenamefont {Cadore}, \citenamefont {De~Fazio},
  \citenamefont {Pingault}, \citenamefont {Mostaani}, \citenamefont {Li},
  \citenamefont {Chen}, \citenamefont {Watanabe}, \citenamefont {Taniguchi},
  \citenamefont {Tongay}, \citenamefont {Wang}, \citenamefont {Ferrari},\ and\
  \citenamefont {Atat{\"u}re}}]{Barbone2018}%
  \BibitemOpen
  \bibfield  {author} {\bibinfo {author} {\bibfnamefont {M.}~\bibnamefont
  {Barbone}}, \bibinfo {author} {\bibfnamefont {A.~R.-P.}\ \bibnamefont
  {Montblanch}}, \bibinfo {author} {\bibfnamefont {D.~M.}\ \bibnamefont
  {Kara}}, \bibinfo {author} {\bibfnamefont {C.}~\bibnamefont
  {Palacios-Berraquero}}, \bibinfo {author} {\bibfnamefont {A.~R.}\
  \bibnamefont {Cadore}}, \bibinfo {author} {\bibfnamefont {D.}~\bibnamefont
  {De~Fazio}}, \bibinfo {author} {\bibfnamefont {B.}~\bibnamefont {Pingault}},
  \bibinfo {author} {\bibfnamefont {E.}~\bibnamefont {Mostaani}}, \bibinfo
  {author} {\bibfnamefont {H.}~\bibnamefont {Li}}, \bibinfo {author}
  {\bibfnamefont {B.}~\bibnamefont {Chen}}, \bibinfo {author} {\bibfnamefont
  {K.}~\bibnamefont {Watanabe}}, \bibinfo {author} {\bibfnamefont
  {T.}~\bibnamefont {Taniguchi}}, \bibinfo {author} {\bibfnamefont
  {S.}~\bibnamefont {Tongay}}, \bibinfo {author} {\bibfnamefont
  {G.}~\bibnamefont {Wang}}, \bibinfo {author} {\bibfnamefont {A.~C.}\
  \bibnamefont {Ferrari}}, \ and\ \bibinfo {author} {\bibfnamefont
  {M.}~\bibnamefont {Atat{\"u}re}},\ }\href {\doibase
  10.1038/s41467-018-05632-4} {\bibfield  {journal} {\bibinfo  {journal}
  {Nature Communications}\ }\textbf {\bibinfo {volume} {9}},\ \bibinfo {pages}
  {3721} (\bibinfo {year} {2018})}\BibitemShut {NoStop}%
\bibitem [{\citenamefont {Qu}\ \emph {et~al.}(2019)\citenamefont {Qu},
  \citenamefont {Braganca}, \citenamefont {Vasconcelos}, \citenamefont {Liu},
  \citenamefont {Xie},\ and\ \citenamefont {Zeng}}]{10.1088/2053-1583/ab2cf7}%
  \BibitemOpen
  \bibfield  {author} {\bibinfo {author} {\bibfnamefont {F.}~\bibnamefont
  {Qu}}, \bibinfo {author} {\bibfnamefont {H.}~\bibnamefont {Braganca}},
  \bibinfo {author} {\bibfnamefont {R.}~\bibnamefont {Vasconcelos}}, \bibinfo
  {author} {\bibfnamefont {F.}~\bibnamefont {Liu}}, \bibinfo {author}
  {\bibfnamefont {S.~J.}\ \bibnamefont {Xie}}, \ and\ \bibinfo {author}
  {\bibfnamefont {H.}~\bibnamefont {Zeng}},\ }\href
  {https://doi.org/10.1088%2F2053-1583%2Fab2cf7} {\bibfield  {journal}
  {\bibinfo  {journal} {2D Materials}\ }\textbf {\bibinfo {volume} {6}},\
  \bibinfo {pages} {045014} (\bibinfo {year} {2019})}\BibitemShut {NoStop}%
\bibitem [{\citenamefont {Cao}\ \emph {et~al.}(2016)\citenamefont {Cao},
  \citenamefont {Tang}, \citenamefont {Sun}, \citenamefont {Peng},
  \citenamefont {Gao}, \citenamefont {Zhao}, \citenamefont {Qian},
  \citenamefont {Sun}, \citenamefont {Ali}, \citenamefont {Shao}, \citenamefont
  {Wu}, \citenamefont {Song}, \citenamefont {Williams}, \citenamefont {Sheng},
  \citenamefont {Jin},\ and\ \citenamefont
  {Xu}}]{http://link.springer.com/article/10.1007/s12274-015-0910-z}%
  \BibitemOpen
  \bibfield  {author} {\bibinfo {author} {\bibfnamefont {S.}~\bibnamefont
  {Cao}}, \bibinfo {author} {\bibfnamefont {J.}~\bibnamefont {Tang}}, \bibinfo
  {author} {\bibfnamefont {Y.}~\bibnamefont {Sun}}, \bibinfo {author}
  {\bibfnamefont {K.}~\bibnamefont {Peng}}, \bibinfo {author} {\bibfnamefont
  {Y.}~\bibnamefont {Gao}}, \bibinfo {author} {\bibfnamefont {Y.}~\bibnamefont
  {Zhao}}, \bibinfo {author} {\bibfnamefont {C.}~\bibnamefont {Qian}}, \bibinfo
  {author} {\bibfnamefont {S.}~\bibnamefont {Sun}}, \bibinfo {author}
  {\bibfnamefont {H.}~\bibnamefont {Ali}}, \bibinfo {author} {\bibfnamefont
  {Y.}~\bibnamefont {Shao}}, \bibinfo {author} {\bibfnamefont {S.}~\bibnamefont
  {Wu}}, \bibinfo {author} {\bibfnamefont {F.}~\bibnamefont {Song}}, \bibinfo
  {author} {\bibfnamefont {D.~A.}\ \bibnamefont {Williams}}, \bibinfo {author}
  {\bibfnamefont {W.}~\bibnamefont {Sheng}}, \bibinfo {author} {\bibfnamefont
  {K.}~\bibnamefont {Jin}}, \ and\ \bibinfo {author} {\bibfnamefont
  {X.}~\bibnamefont {Xu}},\ }\href {\doibase 10.1007/s12274-015-0910-z}
  {\bibfield  {journal} {\bibinfo  {journal} {Nano Res.}\ }\textbf {\bibinfo
  {volume} {9}},\ \bibinfo {pages} {306} (\bibinfo {year} {2016})}\BibitemShut
  {NoStop}%
\bibitem [{\citenamefont {Chen}\ \emph {et~al.}(2016)\citenamefont {Chen},
  \citenamefont {Xing}, \citenamefont {Zhu}, \citenamefont {Zha}, \citenamefont
  {Niu}, \citenamefont {Guo},\ and\ \citenamefont
  {Shao}}]{https://www.doi.org/10.1063/1.4948330}%
  \BibitemOpen
  \bibfield  {author} {\bibinfo {author} {\bibfnamefont {X.}~\bibnamefont
  {Chen}}, \bibinfo {author} {\bibfnamefont {J.}~\bibnamefont {Xing}}, \bibinfo
  {author} {\bibfnamefont {L.}~\bibnamefont {Zhu}}, \bibinfo {author}
  {\bibfnamefont {F.~X.}\ \bibnamefont {Zha}}, \bibinfo {author} {\bibfnamefont
  {Z.}~\bibnamefont {Niu}}, \bibinfo {author} {\bibfnamefont {S.}~\bibnamefont
  {Guo}}, \ and\ \bibinfo {author} {\bibfnamefont {J.}~\bibnamefont {Shao}},\
  }\href {https://www.doi.org/10.1063/1.4948330} {\bibfield  {journal}
  {\bibinfo  {journal} {J. Appl. Phys.}\ }\textbf {\bibinfo {volume} {119}},\
  \bibinfo {pages} {175301} (\bibinfo {year} {2016})}\BibitemShut {NoStop}%
\bibitem [{\citenamefont {Tang}\ and\ \citenamefont {Xu}(2018)}]{Tang27804}%
  \BibitemOpen
  \bibfield  {author} {\bibinfo {author} {\bibfnamefont {J.}~\bibnamefont
  {Tang}}\ and\ \bibinfo {author} {\bibfnamefont {X.-L.}\ \bibnamefont {Xu}},\
  }\href {\doibase 10.1088/1674-1056/27/2/027804} {\bibfield  {journal}
  {\bibinfo  {journal} {Chinese Physics B}\ }\textbf {\bibinfo {volume} {27}},\
  \bibinfo {eid} {27804} (\bibinfo {year} {2018})}\BibitemShut {NoStop}%
\bibitem [{\citenamefont {Hou}\ \emph {et~al.}(1991)\citenamefont {Hou},
  \citenamefont {Staguhn}, \citenamefont {Takeyama}, \citenamefont {Miura},
  \citenamefont {Segawa}, \citenamefont {Aoyagi},\ and\ \citenamefont
  {Namba}}]{https://www.doi.org/10.1103/PhysRevB.43.4152}%
  \BibitemOpen
  \bibfield  {author} {\bibinfo {author} {\bibfnamefont {H.~Q.}\ \bibnamefont
  {Hou}}, \bibinfo {author} {\bibfnamefont {W.}~\bibnamefont {Staguhn}},
  \bibinfo {author} {\bibfnamefont {S.}~\bibnamefont {Takeyama}}, \bibinfo
  {author} {\bibfnamefont {N.}~\bibnamefont {Miura}}, \bibinfo {author}
  {\bibfnamefont {Y.}~\bibnamefont {Segawa}}, \bibinfo {author} {\bibfnamefont
  {Y.}~\bibnamefont {Aoyagi}}, \ and\ \bibinfo {author} {\bibfnamefont
  {S.}~\bibnamefont {Namba}},\ }\href {\doibase 10.1103/PhysRevB.43.4152}
  {\bibfield  {journal} {\bibinfo  {journal} {Phys. Rev. B}\ }\textbf {\bibinfo
  {volume} {43}},\ \bibinfo {pages} {4152} (\bibinfo {year}
  {1991})}\BibitemShut {NoStop}%
\bibitem [{\citenamefont {Kim}\ \emph {et~al.}(2003)\citenamefont {Kim},
  \citenamefont {Fisher}, \citenamefont {Eisler},\ and\ \citenamefont
  {Bawendi}}]{https://www.doi.org/10.1021/ja0361749}%
  \BibitemOpen
  \bibfield  {author} {\bibinfo {author} {\bibfnamefont {S.}~\bibnamefont
  {Kim}}, \bibinfo {author} {\bibfnamefont {B.}~\bibnamefont {Fisher}},
  \bibinfo {author} {\bibfnamefont {H.~J.}\ \bibnamefont {Eisler}}, \ and\
  \bibinfo {author} {\bibfnamefont {M.}~\bibnamefont {Bawendi}},\ }\href
  {\doibase 10.1021/ja0361749} {\bibfield  {journal} {\bibinfo  {journal} {J.
  Am. Chem. Soc.}\ }\textbf {\bibinfo {volume} {125}},\ \bibinfo {pages}
  {11466} (\bibinfo {year} {2003})}\BibitemShut {NoStop}%
\bibitem [{\citenamefont
  {Kamimura}(1986)}]{https://doi.org/10.1016/0038-1098(86)90573-9}%
  \BibitemOpen
  \bibfield  {author} {\bibinfo {author} {\bibfnamefont {H.}~\bibnamefont
  {Kamimura}},\ }\href {\doibase 10.1016/0038-1098(86)90573-9} {\bibfield
  {journal} {\bibinfo  {journal} {Solid State Commun.}\ }\textbf {\bibinfo
  {volume} {59}},\ \bibinfo {pages} {405} (\bibinfo {year} {1986})}\BibitemShut
  {NoStop}%
\bibitem [{\citenamefont {Li}\ \emph {et~al.}(2014)\citenamefont {Li},
  \citenamefont {Chernikov}, \citenamefont {Zhang}, \citenamefont {Rigosi},
  \citenamefont {Hill}, \citenamefont {van~der Zande}, \citenamefont {Chenet},
  \citenamefont {Shih}, \citenamefont {Hone},\ and\ \citenamefont
  {Heinz}}]{http://dx.doi.org/10.1103/PhysRevB.90.205422}%
  \BibitemOpen
  \bibfield  {author} {\bibinfo {author} {\bibfnamefont {Y.}~\bibnamefont
  {Li}}, \bibinfo {author} {\bibfnamefont {A.}~\bibnamefont {Chernikov}},
  \bibinfo {author} {\bibfnamefont {X.}~\bibnamefont {Zhang}}, \bibinfo
  {author} {\bibfnamefont {A.}~\bibnamefont {Rigosi}}, \bibinfo {author}
  {\bibfnamefont {H.~M.}\ \bibnamefont {Hill}}, \bibinfo {author}
  {\bibfnamefont {A.~M.}\ \bibnamefont {van~der Zande}}, \bibinfo {author}
  {\bibfnamefont {D.~A.}\ \bibnamefont {Chenet}}, \bibinfo {author}
  {\bibfnamefont {E.-M.}\ \bibnamefont {Shih}}, \bibinfo {author}
  {\bibfnamefont {J.}~\bibnamefont {Hone}}, \ and\ \bibinfo {author}
  {\bibfnamefont {T.~F.}\ \bibnamefont {Heinz}},\ }\href {\doibase
  10.1103/PhysRevB.90.205422} {\bibfield  {journal} {\bibinfo  {journal} {Phys.
  Rev. B}\ }\textbf {\bibinfo {volume} {90}},\ \bibinfo {pages} {205422}
  (\bibinfo {year} {2014})}\BibitemShut {NoStop}%
\bibitem [{\citenamefont {Carvalho}\ \emph {et~al.}(2017)\citenamefont
  {Carvalho}, \citenamefont {Wang}, \citenamefont {Mignuzzi}, \citenamefont
  {Roy}, \citenamefont {Terrones}, \citenamefont {Fantini}, \citenamefont
  {Crespi}, \citenamefont {Malard},\ and\ \citenamefont
  {Pimenta}}]{https://www.nature.com/articles/ncomms14670}%
  \BibitemOpen
  \bibfield  {author} {\bibinfo {author} {\bibfnamefont {B.~R.}\ \bibnamefont
  {Carvalho}}, \bibinfo {author} {\bibfnamefont {Y.}~\bibnamefont {Wang}},
  \bibinfo {author} {\bibfnamefont {S.}~\bibnamefont {Mignuzzi}}, \bibinfo
  {author} {\bibfnamefont {D.}~\bibnamefont {Roy}}, \bibinfo {author}
  {\bibfnamefont {M.}~\bibnamefont {Terrones}}, \bibinfo {author}
  {\bibfnamefont {C.}~\bibnamefont {Fantini}}, \bibinfo {author} {\bibfnamefont
  {V.~H.}\ \bibnamefont {Crespi}}, \bibinfo {author} {\bibfnamefont {L.~M.}\
  \bibnamefont {Malard}}, \ and\ \bibinfo {author} {\bibfnamefont {M.~A.}\
  \bibnamefont {Pimenta}},\ }\href {https://doi.org/10.1038/ncomms14670}
  {\bibfield  {journal} {\bibinfo  {journal} {Nat. Commun.}\ }\textbf {\bibinfo
  {volume} {8}},\ \bibinfo {pages} {1} (\bibinfo {year} {2017})}\BibitemShut
  {NoStop}%
\bibitem [{\citenamefont {Koperski}\ \emph {et~al.}(2018)\citenamefont
  {Koperski}, \citenamefont {Molas}, \citenamefont {Arora}, \citenamefont
  {Nogajewski}, \citenamefont {Bartos}, \citenamefont {Wyzula}, \citenamefont
  {Vaclavkova}, \citenamefont {Kossacki},\ and\ \citenamefont
  {Potemski}}]{https://doi.org/10.1088/2053-1583/aae14b}%
  \BibitemOpen
  \bibfield  {author} {\bibinfo {author} {\bibfnamefont {M.}~\bibnamefont
  {Koperski}}, \bibinfo {author} {\bibfnamefont {M.~R.}\ \bibnamefont {Molas}},
  \bibinfo {author} {\bibfnamefont {A.}~\bibnamefont {Arora}}, \bibinfo
  {author} {\bibfnamefont {K.}~\bibnamefont {Nogajewski}}, \bibinfo {author}
  {\bibfnamefont {M.}~\bibnamefont {Bartos}}, \bibinfo {author} {\bibfnamefont
  {J.}~\bibnamefont {Wyzula}}, \bibinfo {author} {\bibfnamefont
  {D.}~\bibnamefont {Vaclavkova}}, \bibinfo {author} {\bibfnamefont
  {P.}~\bibnamefont {Kossacki}}, \ and\ \bibinfo {author} {\bibfnamefont
  {M.}~\bibnamefont {Potemski}},\ }\href {\doibase 10.1088/2053-1583/aae14b}
  {\bibfield  {journal} {\bibinfo  {journal} {2D Mater.}\ }\textbf {\bibinfo
  {volume} {6}},\ \bibinfo {pages} {015001} (\bibinfo {year}
  {2018})}\BibitemShut {NoStop}%
\bibitem [{\citenamefont {Dang}\ \emph {et~al.}(2020)\citenamefont {Dang},
  \citenamefont {Sun}, \citenamefont {Xie}, \citenamefont {Yu}, \citenamefont
  {Peng}, \citenamefont {Qian}, \citenamefont {Wu}, \citenamefont {Song},
  \citenamefont {Yang}, \citenamefont {Xiao}, \citenamefont {Yang},
  \citenamefont {Wang}, \citenamefont {Rafiq}, \citenamefont {Wang},\ and\
  \citenamefont {Xu}}]{https://doi.org/10.1038/s41699-020-0136-0}%
  \BibitemOpen
  \bibfield  {author} {\bibinfo {author} {\bibfnamefont {J.}~\bibnamefont
  {Dang}}, \bibinfo {author} {\bibfnamefont {S.}~\bibnamefont {Sun}}, \bibinfo
  {author} {\bibfnamefont {X.}~\bibnamefont {Xie}}, \bibinfo {author}
  {\bibfnamefont {Y.}~\bibnamefont {Yu}}, \bibinfo {author} {\bibfnamefont
  {K.}~\bibnamefont {Peng}}, \bibinfo {author} {\bibfnamefont {C.}~\bibnamefont
  {Qian}}, \bibinfo {author} {\bibfnamefont {S.}~\bibnamefont {Wu}}, \bibinfo
  {author} {\bibfnamefont {F.}~\bibnamefont {Song}}, \bibinfo {author}
  {\bibfnamefont {J.}~\bibnamefont {Yang}}, \bibinfo {author} {\bibfnamefont
  {S.}~\bibnamefont {Xiao}}, \bibinfo {author} {\bibfnamefont {L.}~\bibnamefont
  {Yang}}, \bibinfo {author} {\bibfnamefont {Y.~W.}\ \bibnamefont {Wang}},
  \bibinfo {author} {\bibfnamefont {M.~A.}\ \bibnamefont {Rafiq}}, \bibinfo
  {author} {\bibfnamefont {C.}~\bibnamefont {Wang}}, \ and\ \bibinfo {author}
  {\bibfnamefont {X.}~\bibnamefont {Xu}},\ }\href
  {https://doi.org/10.1038/s41699-020-0136-0} {\bibfield  {journal} {\bibinfo
  {journal} {npj 2D Mater Appl.}\ }\textbf {\bibinfo {volume} {4}},\ \bibinfo
  {pages} {2} (\bibinfo {year} {2020})}\BibitemShut {NoStop}%
\bibitem [{\citenamefont {Yao}, \citenamefont {Xiao},\ and\ \citenamefont
  {Niu}(2008)}]{ISI:000257289500096}%
  \BibitemOpen
  \bibfield  {author} {\bibinfo {author} {\bibfnamefont {W.}~\bibnamefont
  {Yao}}, \bibinfo {author} {\bibfnamefont {D.}~\bibnamefont {Xiao}}, \ and\
  \bibinfo {author} {\bibfnamefont {Q.}~\bibnamefont {Niu}},\ }\href@noop {}
  {\bibfield  {journal} {\bibinfo  {journal} {{Phys. Rev. B}}\ }\textbf
  {\bibinfo {volume} {{77}}},\ \bibinfo {pages} {{23}} (\bibinfo {year}
  {{2008}})}\BibitemShut {NoStop}%
\bibitem [{\citenamefont {MacNeill}\ \emph {et~al.}(2015)\citenamefont
  {MacNeill}, \citenamefont {Heikes}, \citenamefont {Mak}, \citenamefont
  {Anderson}, \citenamefont {Kormanyos}, \citenamefont {Zolyomi}, \citenamefont
  {Park},\ and\ \citenamefont {Ralph}}]{ISI:000348576100021}%
  \BibitemOpen
  \bibfield  {author} {\bibinfo {author} {\bibfnamefont {D.}~\bibnamefont
  {MacNeill}}, \bibinfo {author} {\bibfnamefont {C.}~\bibnamefont {Heikes}},
  \bibinfo {author} {\bibfnamefont {K.~F.}\ \bibnamefont {Mak}}, \bibinfo
  {author} {\bibfnamefont {Z.}~\bibnamefont {Anderson}}, \bibinfo {author}
  {\bibfnamefont {A.}~\bibnamefont {Kormanyos}}, \bibinfo {author}
  {\bibfnamefont {V.}~\bibnamefont {Zolyomi}}, \bibinfo {author} {\bibfnamefont
  {J.}~\bibnamefont {Park}}, \ and\ \bibinfo {author} {\bibfnamefont {D.~C.}\
  \bibnamefont {Ralph}},\ }\href@noop {} {\bibfield  {journal} {\bibinfo
  {journal} {{Phys. Rev. Lett.}}\ }\textbf {\bibinfo {volume} {{114}}},\
  \bibinfo {pages} {{3}} (\bibinfo {year} {{2015}})}\BibitemShut {NoStop}%
\bibitem [{\citenamefont {Cao}\ \emph {et~al.}(2012)\citenamefont {Cao},
  \citenamefont {Wang}, \citenamefont {Han}, \citenamefont {Ye}, \citenamefont
  {Zhu}, \citenamefont {Shi}, \citenamefont {Niu}, \citenamefont {Tan},
  \citenamefont {Wang}, \citenamefont {Liu},\ and\ \citenamefont
  {Feng}}]{https://www.doi.org/10.1038/ncomms1882}%
  \BibitemOpen
  \bibfield  {author} {\bibinfo {author} {\bibfnamefont {T.}~\bibnamefont
  {Cao}}, \bibinfo {author} {\bibfnamefont {G.}~\bibnamefont {Wang}}, \bibinfo
  {author} {\bibfnamefont {W.}~\bibnamefont {Han}}, \bibinfo {author}
  {\bibfnamefont {H.}~\bibnamefont {Ye}}, \bibinfo {author} {\bibfnamefont
  {C.}~\bibnamefont {Zhu}}, \bibinfo {author} {\bibfnamefont {J.}~\bibnamefont
  {Shi}}, \bibinfo {author} {\bibfnamefont {Q.}~\bibnamefont {Niu}}, \bibinfo
  {author} {\bibfnamefont {P.}~\bibnamefont {Tan}}, \bibinfo {author}
  {\bibfnamefont {E.}~\bibnamefont {Wang}}, \bibinfo {author} {\bibfnamefont
  {B.}~\bibnamefont {Liu}}, \ and\ \bibinfo {author} {\bibfnamefont
  {J.}~\bibnamefont {Feng}},\ }\href {https://www.doi.org/10.1038/ncomms1882}
  {\bibfield  {journal} {\bibinfo  {journal} {Nat. Commun.}\ }\textbf {\bibinfo
  {volume} {3}},\ \bibinfo {pages} {887} (\bibinfo {year} {2012})}\BibitemShut
  {NoStop}%
\bibitem [{\citenamefont {Nagler}\ \emph {et~al.}(2017)\citenamefont {Nagler},
  \citenamefont {Ballottin}, \citenamefont {Mitioglu}, \citenamefont
  {Mooshammer}, \citenamefont {Paradiso}, \citenamefont {Strunk}, \citenamefont
  {Huber}, \citenamefont {Chernikov}, \citenamefont {Christianen},
  \citenamefont {Sch\"uller},\ and\ \citenamefont
  {Korn}}]{https://doi.org/10.1038/s41467-017-01748-1}%
  \BibitemOpen
  \bibfield  {author} {\bibinfo {author} {\bibfnamefont {P.}~\bibnamefont
  {Nagler}}, \bibinfo {author} {\bibfnamefont {M.~V.}\ \bibnamefont
  {Ballottin}}, \bibinfo {author} {\bibfnamefont {A.~A.}\ \bibnamefont
  {Mitioglu}}, \bibinfo {author} {\bibfnamefont {F.}~\bibnamefont
  {Mooshammer}}, \bibinfo {author} {\bibfnamefont {N.}~\bibnamefont
  {Paradiso}}, \bibinfo {author} {\bibfnamefont {C.}~\bibnamefont {Strunk}},
  \bibinfo {author} {\bibfnamefont {R.}~\bibnamefont {Huber}}, \bibinfo
  {author} {\bibfnamefont {A.}~\bibnamefont {Chernikov}}, \bibinfo {author}
  {\bibfnamefont {P.~C.~M.}\ \bibnamefont {Christianen}}, \bibinfo {author}
  {\bibfnamefont {C.}~\bibnamefont {Sch\"uller}}, \ and\ \bibinfo {author}
  {\bibfnamefont {T.}~\bibnamefont {Korn}},\ }\href {\doibase
  10.1038/s41467-017-01748-1} {\bibfield  {journal} {\bibinfo  {journal} {Nat.
  Commun.}\ }\textbf {\bibinfo {volume} {8}},\ \bibinfo {pages} {1551}
  (\bibinfo {year} {2017})}\BibitemShut {NoStop}%
\bibitem [{\citenamefont {Stier}\ \emph {et~al.}(2016)\citenamefont {Stier},
  \citenamefont {McCreary}, \citenamefont {Jonker}, \citenamefont {Kono},\ and\
  \citenamefont {Crooker}}]{http://dx.doi.org/10.1038/ncomms10643}%
  \BibitemOpen
  \bibfield  {author} {\bibinfo {author} {\bibfnamefont {A.~V.}\ \bibnamefont
  {Stier}}, \bibinfo {author} {\bibfnamefont {K.~M.}\ \bibnamefont {McCreary}},
  \bibinfo {author} {\bibfnamefont {B.~T.}\ \bibnamefont {Jonker}}, \bibinfo
  {author} {\bibfnamefont {J.}~\bibnamefont {Kono}}, \ and\ \bibinfo {author}
  {\bibfnamefont {S.~A.}\ \bibnamefont {Crooker}},\ }\href {\doibase
  10.1038/ncomms10643} {\bibfield  {journal} {\bibinfo  {journal} {Nat.
  Commun.}\ }\textbf {\bibinfo {volume} {7}},\ \bibinfo {pages} {10643}
  (\bibinfo {year} {2016})}\BibitemShut {NoStop}%
\bibitem [{\citenamefont {Plechinger}\ \emph {et~al.}(2016)\citenamefont
  {Plechinger}, \citenamefont {Nagler}, \citenamefont {Arora}, \citenamefont
  {Granados~del Águila}, \citenamefont {Ballottin}, \citenamefont {Frank},
  \citenamefont {Steinleitner}, \citenamefont {Gmitra}, \citenamefont {Fabian},
  \citenamefont {Christianen}, \citenamefont {Bratschitsch}, \citenamefont
  {Schüller},\ and\ \citenamefont
  {Korn}}]{http://dx.doi.org/10.1021/acs.nanolett.6b04171}%
  \BibitemOpen
  \bibfield  {author} {\bibinfo {author} {\bibfnamefont {G.}~\bibnamefont
  {Plechinger}}, \bibinfo {author} {\bibfnamefont {P.}~\bibnamefont {Nagler}},
  \bibinfo {author} {\bibfnamefont {A.}~\bibnamefont {Arora}}, \bibinfo
  {author} {\bibfnamefont {A.}~\bibnamefont {Granados~del Águila}}, \bibinfo
  {author} {\bibfnamefont {M.~V.}\ \bibnamefont {Ballottin}}, \bibinfo {author}
  {\bibfnamefont {T.}~\bibnamefont {Frank}}, \bibinfo {author} {\bibfnamefont
  {P.}~\bibnamefont {Steinleitner}}, \bibinfo {author} {\bibfnamefont
  {M.}~\bibnamefont {Gmitra}}, \bibinfo {author} {\bibfnamefont
  {J.}~\bibnamefont {Fabian}}, \bibinfo {author} {\bibfnamefont {P.~C.~M.}\
  \bibnamefont {Christianen}}, \bibinfo {author} {\bibfnamefont
  {R.}~\bibnamefont {Bratschitsch}}, \bibinfo {author} {\bibfnamefont
  {C.}~\bibnamefont {Schüller}}, \ and\ \bibinfo {author} {\bibfnamefont
  {T.}~\bibnamefont {Korn}},\ }\href {\doibase 10.1021/acs.nanolett.6b04171}
  {\bibfield  {journal} {\bibinfo  {journal} {Nano Lett.}\ }\textbf {\bibinfo
  {volume} {16}},\ \bibinfo {pages} {7899} (\bibinfo {year}
  {2016})}\BibitemShut {NoStop}%
\bibitem [{\citenamefont {Lyons}\ \emph {et~al.}(2019)\citenamefont {Lyons},
  \citenamefont {Dufferwiel}, \citenamefont {Brooks}, \citenamefont {Withers},
  \citenamefont {Taniguchi}, \citenamefont {Watanabe}, \citenamefont
  {Novoselov}, \citenamefont {Burkard},\ and\ \citenamefont
  {Tartakovskii}}]{https://doi.org/10.1038/s41467-019-10228-7}%
  \BibitemOpen
  \bibfield  {author} {\bibinfo {author} {\bibfnamefont {T.}~\bibnamefont
  {Lyons}}, \bibinfo {author} {\bibfnamefont {S.}~\bibnamefont {Dufferwiel}},
  \bibinfo {author} {\bibfnamefont {M.}~\bibnamefont {Brooks}}, \bibinfo
  {author} {\bibfnamefont {F.}~\bibnamefont {Withers}}, \bibinfo {author}
  {\bibfnamefont {T.}~\bibnamefont {Taniguchi}}, \bibinfo {author}
  {\bibfnamefont {K.}~\bibnamefont {Watanabe}}, \bibinfo {author}
  {\bibfnamefont {K.}~\bibnamefont {Novoselov}}, \bibinfo {author}
  {\bibfnamefont {G.}~\bibnamefont {Burkard}}, \ and\ \bibinfo {author}
  {\bibfnamefont {A.}~\bibnamefont {Tartakovskii}},\ }\href
  {https://doi.org/10.1038/s41467-019-10228-7} {\bibfield  {journal} {\bibinfo
  {journal} {Nat. Commun.}\ }\textbf {\bibinfo {volume} {10}},\ \bibinfo
  {pages} {2330} (\bibinfo {year} {2019})}\BibitemShut {NoStop}%
\bibitem [{\citenamefont
  {Bl\"ochl}(1994)}]{https://link.aps.org/doi/10.1103/PhysRevB.50.17953}%
  \BibitemOpen
  \bibfield  {author} {\bibinfo {author} {\bibfnamefont {P.~E.}\ \bibnamefont
  {Bl\"ochl}},\ }\href {\doibase 10.1103/PhysRevB.50.17953} {\bibfield
  {journal} {\bibinfo  {journal} {Phys. Rev. B}\ }\textbf {\bibinfo {volume}
  {50}},\ \bibinfo {pages} {17953} (\bibinfo {year} {1994})}\BibitemShut
  {NoStop}%
\bibitem [{\citenamefont {Jain}\ \emph {et~al.}(2013)\citenamefont {Jain},
  \citenamefont {Ong}, \citenamefont {Hautier}, \citenamefont {Chen},
  \citenamefont {Richards}, \citenamefont {Dacek}, \citenamefont {Cholia},
  \citenamefont {Gunter}, \citenamefont {Skinner}, \citenamefont {Ceder},\ and\
  \citenamefont {Persson}}]{materialsproject}%
  \BibitemOpen
  \bibfield  {author} {\bibinfo {author} {\bibfnamefont {A.}~\bibnamefont
  {Jain}}, \bibinfo {author} {\bibfnamefont {S.~P.}\ \bibnamefont {Ong}},
  \bibinfo {author} {\bibfnamefont {G.}~\bibnamefont {Hautier}}, \bibinfo
  {author} {\bibfnamefont {W.}~\bibnamefont {Chen}}, \bibinfo {author}
  {\bibfnamefont {W.~D.}\ \bibnamefont {Richards}}, \bibinfo {author}
  {\bibfnamefont {S.}~\bibnamefont {Dacek}}, \bibinfo {author} {\bibfnamefont
  {S.}~\bibnamefont {Cholia}}, \bibinfo {author} {\bibfnamefont
  {D.}~\bibnamefont {Gunter}}, \bibinfo {author} {\bibfnamefont
  {D.}~\bibnamefont {Skinner}}, \bibinfo {author} {\bibfnamefont
  {G.}~\bibnamefont {Ceder}}, \ and\ \bibinfo {author} {\bibfnamefont {K.~A.}\
  \bibnamefont {Persson}},\ }\href {\doibase 10.1063/1.4812323} {\bibfield
  {journal} {\bibinfo  {journal} {APL Mater.}\ }\textbf {\bibinfo {volume}
  {1}},\ \bibinfo {pages} {011002} (\bibinfo {year} {2013})}\BibitemShut
  {NoStop}%
\bibitem [{\citenamefont {Sun}\ \emph {et~al.}(2019)\citenamefont {Sun},
  \citenamefont {Yu}, \citenamefont {Dang}, \citenamefont {Peng}, \citenamefont
  {Xie}, \citenamefont {Song}, \citenamefont {Qian}, \citenamefont {Wu},
  \citenamefont {Ali}, \citenamefont {Tang}, \citenamefont {Yang},
  \citenamefont {Xiao}, \citenamefont {Tian}, \citenamefont {Wang},
  \citenamefont {Shan}, \citenamefont {Rafiq}, \citenamefont {Wang},\ and\
  \citenamefont {Xu}}]{https://aip.scitation.org/doi/10.1063/1.5087440}%
  \BibitemOpen
  \bibfield  {author} {\bibinfo {author} {\bibfnamefont {S.}~\bibnamefont
  {Sun}}, \bibinfo {author} {\bibfnamefont {Y.}~\bibnamefont {Yu}}, \bibinfo
  {author} {\bibfnamefont {J.}~\bibnamefont {Dang}}, \bibinfo {author}
  {\bibfnamefont {K.}~\bibnamefont {Peng}}, \bibinfo {author} {\bibfnamefont
  {X.}~\bibnamefont {Xie}}, \bibinfo {author} {\bibfnamefont {F.}~\bibnamefont
  {Song}}, \bibinfo {author} {\bibfnamefont {C.}~\bibnamefont {Qian}}, \bibinfo
  {author} {\bibfnamefont {S.}~\bibnamefont {Wu}}, \bibinfo {author}
  {\bibfnamefont {H.}~\bibnamefont {Ali}}, \bibinfo {author} {\bibfnamefont
  {J.}~\bibnamefont {Tang}}, \bibinfo {author} {\bibfnamefont {J.}~\bibnamefont
  {Yang}}, \bibinfo {author} {\bibfnamefont {S.}~\bibnamefont {Xiao}}, \bibinfo
  {author} {\bibfnamefont {S.}~\bibnamefont {Tian}}, \bibinfo {author}
  {\bibfnamefont {M.}~\bibnamefont {Wang}}, \bibinfo {author} {\bibfnamefont
  {X.}~\bibnamefont {Shan}}, \bibinfo {author} {\bibfnamefont {M.~A.}\
  \bibnamefont {Rafiq}}, \bibinfo {author} {\bibfnamefont {C.}~\bibnamefont
  {Wang}}, \ and\ \bibinfo {author} {\bibfnamefont {X.}~\bibnamefont {Xu}},\
  }\href {https://aip.scitation.org/doi/10.1063/1.5087440} {\bibfield
  {journal} {\bibinfo  {journal} {Appl. Phys. Lett.}\ }\textbf {\bibinfo
  {volume} {114}},\ \bibinfo {pages} {113104} (\bibinfo {year}
  {2019})}\BibitemShut {NoStop}%
\bibitem [{\citenamefont {He}\ \emph {et~al.}(2015)\citenamefont {He},
  \citenamefont {Clark}, \citenamefont {Schaibley}, \citenamefont {He},
  \citenamefont {Chen}, \citenamefont {Wei}, \citenamefont {Ding},
  \citenamefont {Zhang}, \citenamefont {Yao}, \citenamefont {Xu}, \citenamefont
  {Lu},\ and\ \citenamefont {Pan}}]{https://doi.org/10.1038/nnano.2015.75}%
  \BibitemOpen
  \bibfield  {author} {\bibinfo {author} {\bibfnamefont {Y.-M.}\ \bibnamefont
  {He}}, \bibinfo {author} {\bibfnamefont {G.}~\bibnamefont {Clark}}, \bibinfo
  {author} {\bibfnamefont {J.~R.}\ \bibnamefont {Schaibley}}, \bibinfo {author}
  {\bibfnamefont {Y.}~\bibnamefont {He}}, \bibinfo {author} {\bibfnamefont
  {M.-C.}\ \bibnamefont {Chen}}, \bibinfo {author} {\bibfnamefont {Y.-J.}\
  \bibnamefont {Wei}}, \bibinfo {author} {\bibfnamefont {X.}~\bibnamefont
  {Ding}}, \bibinfo {author} {\bibfnamefont {Q.}~\bibnamefont {Zhang}},
  \bibinfo {author} {\bibfnamefont {W.}~\bibnamefont {Yao}}, \bibinfo {author}
  {\bibfnamefont {X.}~\bibnamefont {Xu}}, \bibinfo {author} {\bibfnamefont
  {C.-Y.}\ \bibnamefont {Lu}}, \ and\ \bibinfo {author} {\bibfnamefont {J.-W.}\
  \bibnamefont {Pan}},\ }\href {\doibase 10.1038/nnano.2015.75} {\bibfield
  {journal} {\bibinfo  {journal} {Nat. Nanotech.}\ }\textbf {\bibinfo {volume}
  {10}},\ \bibinfo {pages} {497} (\bibinfo {year} {2015})}\BibitemShut
  {NoStop}%
\bibitem [{\citenamefont {Srivastava}\ \emph
  {et~al.}(2015{\natexlab{b}})\citenamefont {Srivastava}, \citenamefont
  {Sidler}, \citenamefont {Allain}, \citenamefont {Lembke}, \citenamefont
  {Kis},\ and\ \citenamefont
  {Imamoglu}}]{https://doi.org/10.1038/nnano.2015.60}%
  \BibitemOpen
  \bibfield  {author} {\bibinfo {author} {\bibfnamefont {A.}~\bibnamefont
  {Srivastava}}, \bibinfo {author} {\bibfnamefont {M.}~\bibnamefont {Sidler}},
  \bibinfo {author} {\bibfnamefont {A.~V.}\ \bibnamefont {Allain}}, \bibinfo
  {author} {\bibfnamefont {D.~S.}\ \bibnamefont {Lembke}}, \bibinfo {author}
  {\bibfnamefont {A.}~\bibnamefont {Kis}}, \ and\ \bibinfo {author}
  {\bibfnamefont {A.}~\bibnamefont {Imamoglu}},\ }\href {\doibase
  10.1038/nnano.2015.60} {\bibfield  {journal} {\bibinfo  {journal} {Nature
  Nanotechnology}\ }\textbf {\bibinfo {volume} {10}},\ \bibinfo {pages} {491}
  (\bibinfo {year} {2015}{\natexlab{b}})}\BibitemShut {NoStop}%
\end{thebibliography}
\end{document}